\documentclass[%
 reprint,
 amsmath,amssymb,
 aps,
]{revtex4-1}

\usepackage{mathrsfs}
\usepackage{color}
\usepackage[thicklines]{cancel}

\usepackage{soul}
\setstcolor{red}

\definecolor{mygreen}{rgb}{0, 0.6, 0}

\def\bea{\begin{eqnarray}}
\def\eea{\end{eqnarray}}

\usepackage{graphicx}
\usepackage{dcolumn}
\usepackage{bm}

\begin{document}


\title{Role of particle local curvature in cellular wrapping}

\author{Amir Khosravanizadeh}
\email{khosravani@iasbs.ac.ir}
\affiliation{Department of Physics, Institute for Advanced Studies in Basic Sciences (IASBS), Zanjan 45137-66731, Iran}
\affiliation{Université Paris Cité, CNRS, Institut Jacques Monod, F-75013 Paris, France}
\author{Pierre Sens}
\affiliation{Institut Curie, PSL Research University, CNRS UMR 168; F-75005 Paris, France}
\author{Farshid Mohammad-Rafiee}
\email{farshid@iasbs.ac.ir}
\affiliation{Department of Physics, Institute for Advanced Studies in Basic Sciences (IASBS), Zanjan 45137-66731, Iran}
\affiliation{Research Center for Basic Sciences \& Modern Technologies $(\text{RBST})$, Institute for Advanced Studies in Basic Sciences (IASBS), Zanjan 45137-66731, Iran}

\date{June, 2022}

\begin{abstract}
Cellular uptake through the lipid membranes plays an important role in adsorbing nutrients and fighting infection and can be used for drug delivery and nanomedicine developments. Endocytosis is one of the known pathways of the cellular uptake which associate with elastic deformation of the membrane wrapping around the foreign particle. The deformability of the membrane itself is strongly regulated by the presence of a cortical cytoskeleton placed underneath the membrane. It has been shown that size, shape, and orientation of the particles influence on their entry into the cell. Here, we study the role of particle local curvature in the cellular uptake by investigating the wrapping of an elastic membrane around a long cylindrical object with an elliptical cross section. The membrane itself is adhered to a substrate mimicking the cytoskeleton. Membrane wrapping proceeds differently whether the initial contact occurs at the particle's highly curved tip (prolate) or along its long side (oblate). We obtain a wrapping phase diagram as a function of the membrane-cytoskeleton and the membrane-target adhesion energy, which includes three distinct regimes of engulfment(unwrapped, partially wrapped, and fully wrapped), separated by two phase transitions. We also provide analytical expressions for the boundary between the different regimes which confirm that the transitions strongly depend on the orientation and aspect ratio of the particle.
\end{abstract}

\pacs{Valid PACS appear here}
\maketitle
\section{Introduction}
Internalization of particles is an essential cellular process by which cells adsorb nutrients and fight infections\cite{lodish}. The interaction of nanoparticles with lipid membranes also provides a broad range of potential applications in biomedical fields such as chemotherapy, bioimaging, biosensing, and drug and gene delivery\cite{xia2008nanomaterials,weissleder2006molecular,nel2009understanding,allen2004drug,whitehead2009knocking,peer2007nanocarriers}. Receptor-mediated endocytosis and phagocytosis are two prominent pathways of cellular uptake\cite{lodish}. These complex cellular processes involve energy consumption and cytoskeleton rearrangement\cite{conner2003regulated,swanson2008shaping}. However, the mechanics and dynamics of internalization can be studied by simplified theoretical and computational approaches to provide a quantitative physical model of endocytosis\cite{gao2005mechanics,vacha2011receptor,li2012molecular,yi2011cellular,yuan2010variable,reynwar2007aggregation,yang2010computer}. In most of these approaches, the wrapping process is considered as the minimization of the total system energy, including the elastic energy of the membrane and adhesion energy to the target\cite{deserno2004elastic,agudo2021particle,weikl2003indirect} or a substrate\cite{khosravanizadeh2019wrapping,hashemi2014regulation,yue2013molecular}. Therefore, for the sake of simplicity, the vast majority of previous papers have studied spherical\cite{deserno2004elastic,muller2005interface,irmscher2013method,mirigian2013kinetics} or cylindrical particles\cite{weikl2003indirect,khosravanizadeh2019wrapping,hashemi2014regulation,muller2007balancing}. 

In realistic situations, cells often uptake targets of different shapes\cite{clarke2010curvature,rittig1992coiling,horwitz1984phagocytosis,cureton2009vesicular,ge2010cryo}. For example, cells are capable of interacting with rod-like bacteria or dumbbell-shaped dividing cells. As a result, the subject matter of recent studies turned to the role of particle shape and orientation in the wrapping process\cite{vacha2011receptor,champion2006role,champion2009shape,gratton2008effect,chen2016shape,shi2011cell,chithrani2006determining,dasgupta2014shape,dasgupta2013wrapping,richards2016target,bahrami2013orientational,baranov2021modulation}, but the results remain contentious. Some of these studies shown that the uptake of rod-like particles, with a high aspect ratio, is less likely than that of spherical particles with similar size\cite{chithrani2006determining,chithrani2007elucidating}, but the opposite behavior has also been reported from both experiment and simulations\cite{vacha2011receptor,gratton2008effect}.

\begin{figure*}[ht!]
\centering
\includegraphics[width=1.6\columnwidth]{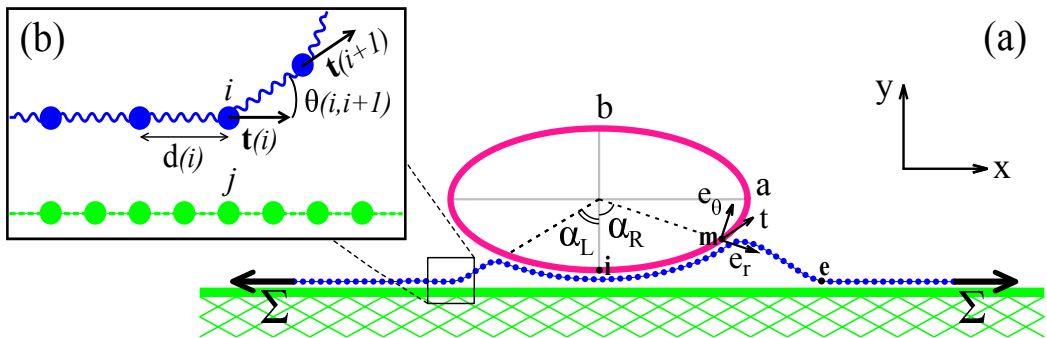}
\caption{(a) The schematic picture of an infinitely long cylindrical particle with elliptical cross section (pink) engulfed by a cellular membrane (blue) attached to a non-deformable cytoskeleton substrate (green). The surface tension of the membrane is denoted by $\Sigma$. The 
ellipse axes in the $x$ and $y$ direction are of length $2a$ and $2b$, respectively.  The wrapping angle at the right and left side of the particle are represented by $\alpha_R$ and $\alpha_L$, respectively. $\hat{t}$ is the tangential unit vector of the ellipse and the radial and azimuthal unit vectors are shown by $\hat{e}_r$, and $\hat{e}_{\theta}$, respectively. (b) a closed view of the bead-spring model of the membrane.}
\label{fig1}
\end{figure*}
In the case of ellipsoidal particles, Bahrami\cite{bahrami2013orientational} analytically found that internalization of these particles depends on their orientation and is more restrictive than spherical ones. The experimental studies of Champion and Mitragori also showed that phagocytosis of ellipsoidal disks depends on  their orientation with respect to the membrane\cite{champion2006role}. Engulfment was easiest when particles contacted the cell through their highly curved tip, and wrapping was generally incomplete when initial contact occurred through the particle's long side. These kind of experiments indicated that beside the shape of particles their orientation needs to be studied, as well. Dasgupta et. al.\cite{dasgupta2014shape,dasgupta2013wrapping} explained that high aspect ratio particles with round tip tend enter to the cell from their long side, while particles with small aspect ratio and a flat tip, enter from the tip. An MD simulation with spherocylindrical particles, showed that when these particles initially have an upright docking position on the membrane, they rotate and first adhere to the membrane from long side. Then they stand up again and complete the engulfment when their long axis is perpendicular to the membrane\cite{huang2013role}.

One of the essential part of cell's structure is an actin network underneath the membrane named cytoskeleton\cite{bray}. The crucial role of the cytoskeleton in the wrapping process should be considered from two aspects. First, it controls the shape of the cell and restrict the deformation of membrane anchored to it, and second, it can generate active forces to push the membrane protrusions around the target. The present model focuses on the first aspect, and considers an inert cytoskeleton on which the membrane adheres. As such, it is also appropriate to study the wrapping of particles by supported membranes adhered on a substrate.

In a previous paper, we considered the effect of cytoskeleton as a fixed substrate under the membrane to study the wrapping of a long cylindrical particle\cite{khosravanizadeh2019wrapping}. In the present paper, we use the same methodology to study the impact of the curvature on the target on the wrapping process. For this, we simulate the cellular wrapping of a long particle with elliptical cross section (Fig. \ref{fig1}). Because of the translational symmetry along the particle axis, this case can be considered as a 2D problem. In one case the particle initially contacts the membrane through its curved tip (prolate state), and in another case from its long side (oblate configuration).

In section 2, we present the details of the coarse grained molecular dynamics model which we used for simulations. In section 3 we summarize the simulation results of oblate and prolate configurations in two phase diagram including three distinct regimes (unwrapped, partially, and fully wrapped) accompanied by two phase transitions. These results show that oblate and prolate ellipses have two different phase diagrams. While wrapping is triggered more easily for the oblate ellipse, the full-wrapped configuration is more favorable for prolate one. The differences in the phase diagrams is explained by the theoretical calculations of transitions between the three regimes. 
\section{Model and Methods}
The simulations presented in this work are based on a highly coarse-grained description of the membrane which we introduced in the previous paper\cite{khosravanizadeh2019wrapping}. This model is appropriate for simulating large scale membranes under finite tension. In addition, because the model is obtained form the discretization of the Helfich energy\cite{helfrich1973elastic}, the results can be easily compared with theoretical calculations.

The cellular membrane (in blue in Fig. \ref{fig1}) is initially adhered to a flat cytoskeleton cortex surface (in green). Owing to adhesion to an external particle (in pink), an infinity long cylinder with elliptical cross section, the membrane deforms and wrap around the target. The ellipse which centered at the origin is characterized as:
\begin{equation}\label{el_cartezian}
\frac{x^2}{a^2}+\frac{y^2}{b^2}=1
\end{equation}
where $2a$ and $2b$ are the length of axes in the $x$ and $y$ direction, respectively. In our simulations, the particle's dimension are fixed to $70 \, \sigma$ and $140 \, \sigma$ (where $\sigma$ is the simulations unit length scale); therefore the aspect
ratio for the oblate case is $D=\frac{a}{b}=2$ and for prolate one $D=1/2$. 

The adhesion energy difference (per unit length) with respect to the initial state can be written as:
\begin{equation}
\Delta E_{ad}=-\omega \int r(\alpha) d \alpha + \omega_s (l_{s,L}+l_{s,R})
\end{equation}
where 
$\omega$ and $\omega_s$ denote the particle-membrane and the cytoskeleton-membrane adhesion energy per unit area, respectively. The membrane wrapping around the target is characterized by the wrapping angle $\alpha=\alpha_R+\alpha_L$, where $\alpha_R$ and $\alpha_L$ are the wrapping angle at the right and left side of the particle, respectively. $l_{s,L}$ and $l_{s,R}$ indicate the left and right membrane contour length detached from the cytoskeleton (the contour length between the points $i$ and $e$ in Fig. \ref{fig1}(a). One should note that in principle the wrapping angle can be asymmetric\cite{khosravanizadeh2019wrapping} and therefore $l_{s,L}$ can be different from $l_{s,R}$. 

The elastic energy of the deformed membrane can be described by the Helfrich energy\cite{helfrich1973elastic} as
\begin{equation}
E_H=\frac{1}{2}\kappa\int_a \left( 2H \right)^{2}da+\Sigma\int_a da,
\end{equation}
where $\kappa$ and $\Sigma$ denote the bending rigidity and the surface tension of the membrane, respectively, and $H$ is the mean curvature of the membrane. Long particles are invariant along their long axis which means a single tangent vector, $\hat{t}(s)$, defined at each point of the membrane, $s$, suffices for fully characterization of its shape. Therefore the total deformation energy per unit length of the particle can be written as
\begin{equation}
\frac{E_H}{L} =\frac{1}{2}\kappa\int_s \left[ \partial_s \hat{t}(s) \right]^{2} ds+\Sigma\int_s ds,
\label{one_dimension}
\end{equation}
where $\partial_s \hat{t}(s)$ indicates differentiation with respect to the arclength $s$ and $L$ is the length of the cylinder. Consequently, this 1D integral can be discretized as a chain of beads which connected to each other by a harmonic spring potential,
\begin{equation}\label{spring_potential}
E_{spring}=\frac{1}{2} \Lambda \sum_{i=1}^{N-1} \left[ {d\left(i\right)-d_0}\right]^{2}.
\end{equation}
where $d\left(i\right)$ is the bond length, $d_0$ is the equilibrium bond length, and $\Lambda$ is the spring's stiffness. The bending energy of the membrane preserves as the following potential between each three connected beads (Fig. \ref{fig1}(b)),
\begin{equation}\label{bending}
E_B=\frac{\kappa}{d_0} \sum_{i=1}^{N-2} \left[1-\cos \left(\theta \left(i,i+1\right) \right)  \right], 
\end{equation}
where $\theta(i,i+1)$ represents the angle between neighboring springs. 
Furthermore, adding a lateral force at the edges of the membrane reproduces the effect of the membrane surface tension. In this paper, the membrane is constructed by $1000$ monomers, is allowed to move only in $x-y$ plane, and the excluded volume interactions between the membrane beads are implemented using the Weeks-Chandler-Andersen potential
\begin{align}\label{WCA}
V_{WCA}\left(r_{ij}\right)=\left\lbrace 
\begin{array}{cc}
4\varepsilon \left[ \left( \frac{\sigma}{r_{ij}}\right) ^{12}-\left( \frac{\sigma}{r_{ij}}\right) ^{6}+\frac{1}{4} \right],  & r_{ij}\leq r_c\\
0, & r_{ij}> r_c,
\end{array} \right.
\end{align}
where $r_c =2^{1/6} \, \sigma$, $r_{ij}$ is the distance between the $i$th and $j$th beads and $\varepsilon$ and $\sigma$ are the unit energy and length scale of the simulation, respectively. The diameter of the membrane's monomers is $R=d_0 \doteq 1\sigma$. 

The cytoskeleton is implemented by $2600$ immobile beads laid underneath the membrane. The elliptical cross section of the particle is also made of $678$ fixed monomers positioned on the top of the membrane. The distance between the monomers forming the cytoskeleton or the ellipse is taken to be $0.5 \, \sigma$ to simplify the reptation of the membrane on the cytoskeleton.

Both the membrane-cytoskeleton and the membrane-particle interactions are modeled with the following potential:
\begin{eqnarray}
V\left(r_{ij}\right) = 
  \begin{cases}
 4\lambda_{k} \left[ \left( \frac{\sigma}{r_{ij}}\right) ^{12}-\left( \frac{\sigma}{r_{ij}}\right) ^{6} \right],  &r_{ij} < r_{c} \\
-\lambda_{k} \cos^{2}\left[\frac{\pi}{2\zeta}\left(r_{ij}-r_{c}\right)\right] , &r_c \leq r_{ij} \leq r_{c}+\zeta \\
0,  &r_{ij} > r_{c}+\zeta,
  \end{cases} 
  \label{eq:dt}
\end{eqnarray}
where $\lambda_k$, in the unit of the energy, denotes the strength of the ligand-receptor interactions ($k=1$ corresponds to the membrane-target, and $k=2$ corresponds to the membrane-cytoskeleton), and $\zeta=0.5 \, \sigma$. The values of the average adhesion energy per unit length $\sigma$ between the membrane and cytoskeleton ($\omega_s$) and between the membrane and the particle ($\omega$) can be tuned by varying $\lambda_1$ and $\lambda_2$. 

Our Molecular Dynamics (MD) simulations were performed at the constant temperature $k_BT=1.0\varepsilon$, with the Langevin thermostat, and using ESPResSo\cite{espresso}. The time step in the Verlet algorithm and the damping constant in the Langevin thermostat were set $\delta t=0.01 \tau_0$ and $\Gamma=\tau^{-1}_0$, respectively, which $\tau_0=\sqrt{\frac{m\sigma^{2}}{\varepsilon}}$ is the MD time scale and $m$ is the monomer mass. 
\begin{figure*}[ht!]
\centering
\includegraphics[width=1.8\columnwidth]{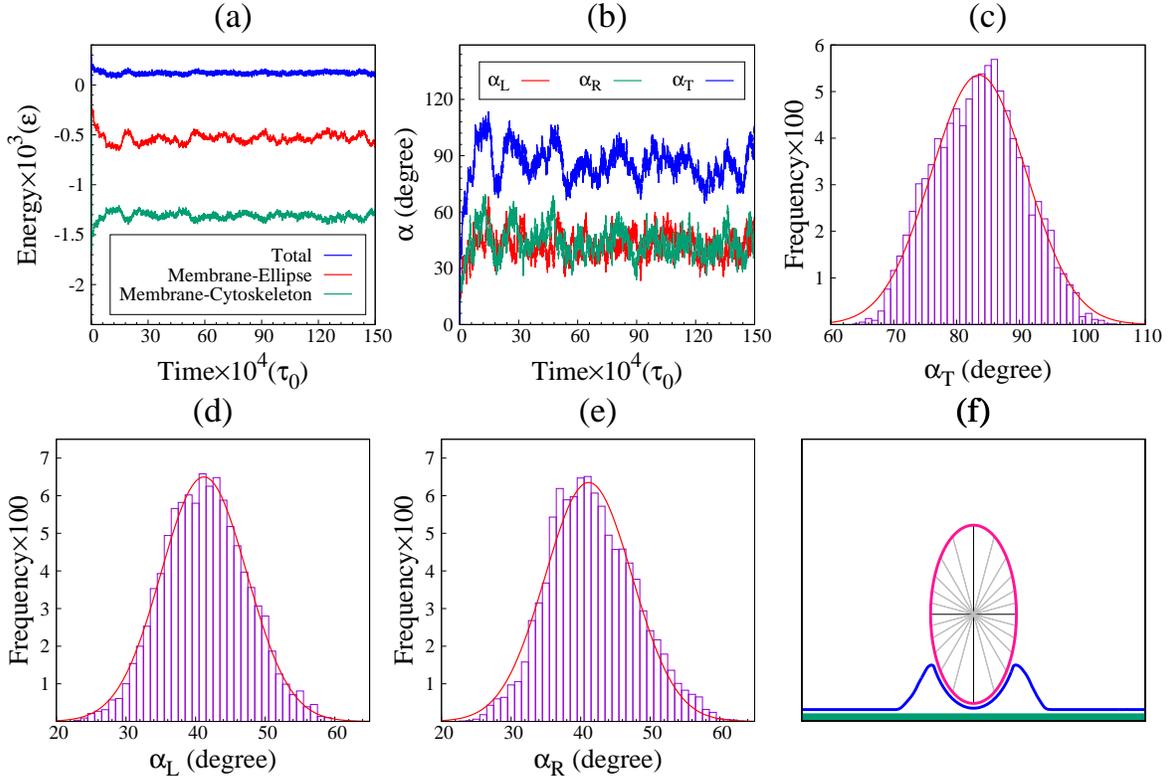}
\caption{The results of a typical simulation of a prolate ellipse corresponding to $\omega_s=2.30 \, \varepsilon/{\sigma}^2$ and $\omega=7.28 \, \varepsilon/{\sigma}^2$. Time evolution of (a) the total energy (blue), membrane-cytoskeleton adhesion energy (green) and membrane-target adhesion energy (red); 
(b) the total (blue), left (red) and right wrapping angles (green). Panels (c) to (e) represent the distribution of the total, the left, and the right wrapping angles in the steady-state, respectively. Panel (f) display a typical snapshot of the membrane conformation in the equilibrium state. More simulation results for both prolate and oblate configurations can be found in Fig. \ref{fig_sup}.}
\label{fig2}
\end{figure*}
\section{Results}
\subsection{Numerical results}
To investigate the role of the particle's local curvature in the wrapping process we study the cases where the ellipse is introduced to the membrane from its narrow tip (prolate configuration) or its long side (oblate configuration). The results of these two configurations can be compared with the results of a cylindrical particle\cite{khosravanizadeh2019wrapping}. 
\begin{figure*}[ht!]
\centering
\includegraphics[width=2\columnwidth]{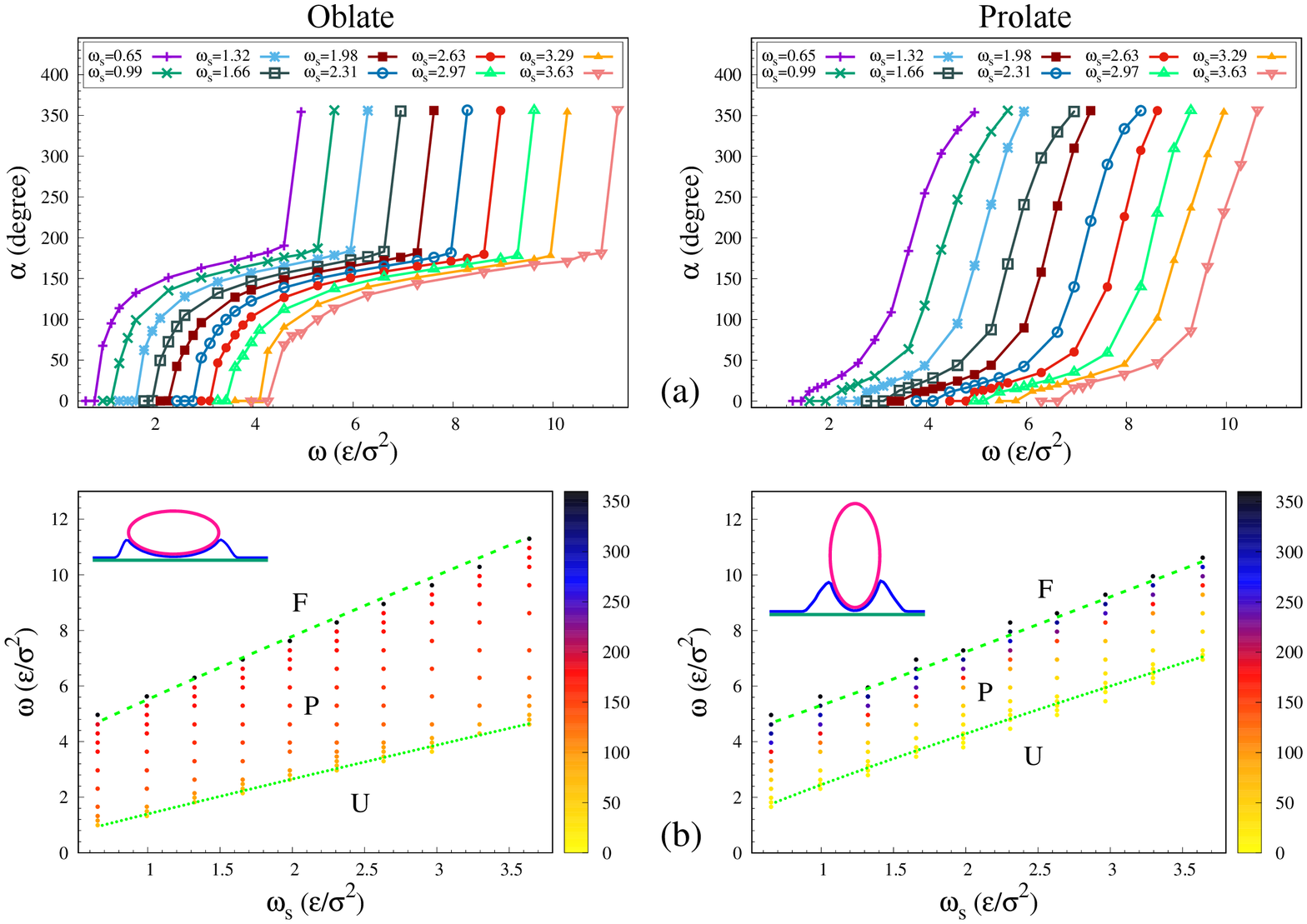}
\caption{(a) Variation of the wrapping angle as a function of $\omega$ for different values of $\omega_s$ in the case of oblate (left) and prolate (right) configuration of the object. (b) Wrapping phase diagrams of the particle in the phase space of $\omega_{s}$ and $\omega$. The color bar shows the extent of wrapping angle (in degrees).
There are three distinct regimes: unwrapped (U), partially wrapped (P), and fully wrapped. The dotted lines corresponds to the transition to a partially wrapped state (Eq. \ref{el_omega-u}) while the dashed lines represent the theoretical value of $\omega$ in transition to the fully wrapped state (Eq. \ref{el_omega-f}).}
\label{fig3}
\end{figure*}

In summary our model contains two membrane elastic parameters (bending rigidity $\kappa$ and surface tension $\Sigma$), two adhesion parameters (cortex and target adhesion energies $\omega_s$ and $\omega$), and one geometrical parameter (aspect ratio $D$). We fixed the values of $\Lambda=5000 \, \varepsilon/\sigma^2$, $\kappa=20 \, \varepsilon$, and $\Sigma=1.5 \, \varepsilon/\sigma^2$, and investigate the wrapping process as a function of the cortex adhesion energy $\omega_s= [0.65-3.63] \, \varepsilon/\sigma^2$ and the target adhesion energy $\omega= [0.65-11.30] \, \varepsilon/\sigma^2$. For this, we change the values of $\lambda_1$ and $\lambda_2$ in the range of $[0.3 - 1.2] \, \varepsilon$ and $[0.3 - 3.5] \, \varepsilon$, respectively. Considering $\sigma=3 \, nm$ as the thickness of a lipid membrane, and $\varepsilon=1 \, k_BT$ the mentioned values corresponds to $\kappa=60 \, k_BT$, $\Sigma= 0.17 \, (k_BT/nm^2)$, $\omega_s =[0,072-0.4]\, (k_BT/nm^2)$, and $\omega = [0,072-1.26]\, (k_BT/nm^2)$.

After a sufficient number of MD steps, the system reaches an equilibrium state where the representing parameters of the system fluctuate around their equilibrium values. The equilibrium state of the system can be specified by looking at the time variation of the energies, and the wrapping angles of the particle. Fig. \ref{fig2} represents the results of a typical simulation of a prolate ellipse corresponding to $\omega_s=2.30 \, \varepsilon/{\sigma}^2$ and $\omega=7.28 \, \varepsilon/{\sigma}^2$ (see also Fig. \ref{fig_sup}). Panel (a) shows the time course of the total system's energy ( blue), the membrane-particle adhesion energy (red), and the membrane-cytoskeleton adhesion energy (green). Panel (b) represents the total (blue), left (red) and right (green) wrapping angles. Panels (c) to (e) represent the distribution of the total wrapping angle, $\alpha$, the left side , $\alpha_L$, and the right side of the particle, $\alpha_R$, when the system is equilibrated.
Corresponding to the panels (c) to (e), these distributions are Gaussian which are in agreement with thermal fluctuation of energies, and wrapping angles in the panels (a) and (b). A typical snapshot of the membrane conformation around the prolate ellipse is displayed in the panel (f).

The  results of the simulations are summarized in Fig. \ref{fig3}. In Fig. \ref{fig3}(a) the equilibrium wrapping angle, $\alpha=\alpha_R+\alpha_L$, is represented as a function of the target adhesion energy, $\omega$, for different values of the cytoskeleton adhesion energy, $\omega_s$, in the prolate and oblate configurations. The phase diagram of the systems as a function of $\omega$ and $\omega_s$ are shown in Fig. \ref{fig3}(b). For small values of $\omega$, the system is in the unwrapped regime ($U$). Beyond a threshold value $\omega_u$, which increase with $\omega_s$, the membrane stably wraps around the target and the equilibrium wrapping angle increases continuously with the target adhesion, $\omega$. This is the partially wrapped regime ($P$) and it is nearly symmetric on the left and right sides. Beyond yet another threshold value, $\omega_f$, which also increases with $\omega_s$, another transition happens and the particle is fully wrapped by the membrane, the fully-wrapped state ($F$). In this regime, the left and right angles can be asymmetric.

Fig. \ref{fig3} shows that the wrapping behavior is strongly influenced by the particle's orientation. The threshold value of the adhesion energy for partial wrapping, $\omega_u$, is smaller in the oblate case than in the prolate case. This means that in the early stages of the cellular wrapping, adhesion is easier for particles with smaller curvatures. In the partially wrapped regime on the other hand, the wrapping angle increases sharply with the adhesion energy $\omega$ in the prolate case, while it remains confined to half wrapping ($\alpha\simeq180^\circ$) for a broad range of $\omega$ in the oblate case. In fact, $\alpha$ is always larger in the oblate case for $\alpha<180^\circ$ and is always larger in the prolate case when $\alpha>180^\circ$. These results are in agreement with the simulations done by Huang et. al \cite{huang2013role}, where they showed that spherocylindrical particles tend to first adhere to the membrane from long side, then they stand up again and complete the engulfment when their long axis is perpendicular to the membrane. Our simulations show that, all parameters being same, the full wrapping threshold values $\omega_f$ is smaller in the prolate case than in the oblate one.
 
It should be noticed that the MD simulations have been used to find the equilibrium conformations of the membrane, which corresponds to the minimum energy of the system. Although we considered a 2D case, a very long particle, our results regarding the role of local curvature on the wrapping process can be expanded to real cases in 3D like phagocytosis, endocytosis, and viral infection. However, these results still can be compared to the semi two dimensional experiments done by Champion et. al \cite{champion2006role}.   
\subsection{Analytical results}
In the following sections, we calculate the analytical expression for the two threshold values $\omega_u$ and $\omega_f$. Here, we focus on the partially wrapped regime and assume that the left and right sides are nearly symmetric, so we can write $\alpha_R=\alpha_L=\theta$ and $l_{s,R}=l_{s,L}=l_s$.
\subsubsection{Unwrapped-partial wrapped (U-P) transition}
The transition from the unwrapped regime to the partial wrapped state ($U-P$ transition) can be described analytically by expanding the total energy of the system for small wrapping angles $\theta \ll 1$. In general, the total energy of the system can be divided in two main parts: the cap (in contact with the target) and the free part (the part which the membrane is neither in contact with the target, nor with the cytoskeleton). At the $U-P$ transition, the wrapping angle is zero, $\theta = 0$. As shown in the appendix (Eq. (\ref{el_curvature})), the particle's local radius of curvature at this point is $r_{eff}=a^2/b=Da$. At the $U-P$ transition, this radius can be considered as the effective radius of a cylinder and be substituted into the analytical expression obtained in \cite{khosravanizadeh2019wrapping} and briefly summarised below. The contributions of the cap and free parts to the energy difference (per unit length of the particle) with respect to the reference state, for which the membrane is fully adhered to the cytoskeleton, are:
\begin{equation}
\Delta E_{cap}=\frac{\kappa \theta}{r_{eff}}+2\Sigma \theta r_{eff} \left( 1- \frac{\sin \theta}{\theta} \right)+2(\omega_s-\omega)r_{eff}\theta,
\label{e_cap}
\end{equation}
and
\begin{equation}
\Delta E_{free}=2\times\int_{0}^{S}ds \left[\frac{\kappa}{2} \dot{\psi}^2+\Sigma \left( 1-\cos\psi \right)+\omega_s \right],
\label{e_free} 
\end{equation}
where $\dot{\psi}=\frac{d \psi}{d s}$, and $S$ represent the total contour length of the membrane in the free segment. For small $\alpha$, the energy difference of the system per unit length of the particle has a general form as
\begin{equation}
 \Delta E = \frac{\kappa}{r_{eff}} \left[ A_1 \theta + A_2 \theta^3 \right], \label{eq:E-small-angle}
\end{equation}
where $A_1$ and $A_2$ are functions of physical parameters of the system and can be found in explicit form in ref. \cite{khosravanizadeh2019wrapping}. The wrapping angle can be found by minimizing Eq. (\ref{eq:E-small-angle}) with respect to $\theta$. We note that when $A_1$ becomes zero, the transition from the unwrapped regime to the partial wrapped regime happens. Using this criterion, we find a threshold for the membrane-target adhesion energy for unwrapped-partial wrapped transition as 
\begin{multline}
\omega_u=\frac{\kappa}{2D^2a^2} \left[  \left( 1+\sqrt{D^2\bar{\omega}_s} \right)^2+ \right. \\
 \left. \frac{4}{9}\left( \left( 1+3\sqrt{D^2\bar{\omega}_s} \right)^{3/2}-1 \right) \right] , 
\label{el_omega-u} 
\end{multline}
where $\bar{\omega}_s$ is defined as $\bar{\omega}_s \equiv \frac{2 \omega_s a^2}{\kappa}$.  This transition is indicated by the dotted lines separating the unwrapping region from the partial wrapping region in Fig. \ref{fig3} (b). 
\begin{figure*}[t]
\centering
\includegraphics[width=2\columnwidth]{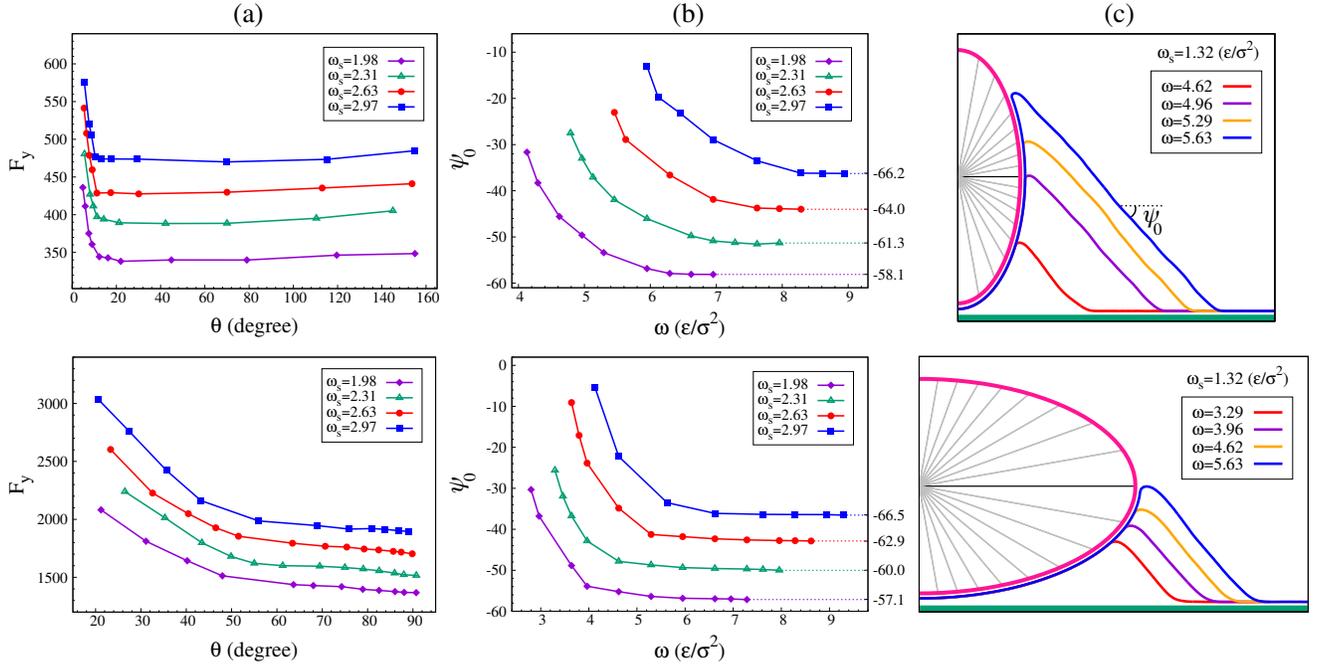}
\caption{Quantification of the wrapping for the two particle orientations: prolate (top) and  oblate (bottom). (a): The vertical component of the acting force on the membrane, $\mathscr{F}_y$, as a function of $\theta$, corresponding to the different values of $\omega$ in the simulation. (b) : The average angle of the free membrane segment (where the membrane is neither attached to the particle nor to the cytoskeleton, see panel (c)). The graphs have been translated along the vertical axis to avoid overlaps. (c): Typical snapshots of the membrane conformation in the the partially wrapped regime.}
\label{fig_comparison}
\end{figure*}
\subsubsection{Partial wrapped-full wrapped (P-F) transition} \label{PFtheory}
In order to understand the transition from the  partial wrapped to the full wrapped state ($P-F$ transition), first we need to derive the force (per unit length) acting on the membrane by calculating energy changes associated to infinitesimal membrane displacements. The generalized force per unit length of the cylinder acting on the membrane, $\vec{\mathscr{F}}$, is given by\cite{hashemi2014regulation,khosravanizadeh2019wrapping}
\begin{equation}
\vec{\mathscr{F}} (s) = \left[ \frac{1}{2} \kappa \dot{\psi}^2 - (\Sigma+\omega_s) \right] \hat{t}(s)+ \kappa \ddot{\psi} \, \hat{n}(s), \label{eq:generalized-force}
\end{equation}
where $\ddot {\psi}$ is $\frac{\partial^2 \psi}{\partial s^2}$. At the equilibrium, the total force acting on each segment of the membrane should be zero, which means that $\vec{\mathscr{F}} (s)$ does not depend on $s$ and is constant. By decomposing $\vec{\mathscr{F}}$ in $x$ and $y$ directions, we can write
\begin{equation}
\frac{\kappa }{2} \dot{\psi}^2 - (\Sigma+\omega_s)= \mathscr{F}_x \cos \psi + \mathscr{F}_y \sin \psi. \label{force-relation1}
\end{equation}
Using Eq. (\ref{force-relation1}) and the boundary conditions at point $e$, $\psi_e=0$ and $\dot{\psi}_e=\sqrt{\frac{2\omega_s}{\kappa}}$\cite{Landau_Elasticity}, the horizontal component of the force can be determined as 
\begin{equation}
\mathscr{F}_x=-\Sigma.
\label{f_x}
\end{equation}
The vertical component of the force can also be derived by applying the boundary condition at point $m$, $\psi_m$ and $\dot{\psi}_m$, in Eq. (\ref{force-relation1})
\begin{equation}
\mathscr{F}_y = \frac{1}{\sin \psi_m} \left[ \frac{\kappa}{2} \dot{\psi}_m^2- \Sigma ( 1 - \cos \psi_m) - \omega_s \right],
\label{f_y}
\end{equation} 
where $\dot{\psi}_m$ can be found as\cite{Landau_Elasticity}
\begin{align}\label{el_psidot_m}
\dot{\psi}_m &=\left[ C(\theta) -\sqrt{2\omega/\kappa} \right]\nonumber \\ &=\left[ \frac{D^2}{a} \left( \frac{D^2+\tan^2\theta}{D^4+\tan^2\theta} \right)^{3/2}-\sqrt{2\omega/\kappa}\right], 
\end{align}
where we replaced $C(\theta)$ from Eq. (\ref{el_curvature}). Fig. \ref{fig_comparison}(c) represents the behavior of $\mathscr{F}_y$ with respect to $\theta$, corresponding to the different values of $\omega$ in the simulation. This force is constant for large enough values of $\theta$ in both prolate and oblate cases. It means that for large enough values of $\theta$ both vertical and horizontal components of the force at the free part of the membrane remain constant. 

Inspection of the membrane shape in Fig. \ref{fig2} and Fig. \ref{fig_comparison}(a) (see also Fig. \ref{fig_sup})
shows that when the wrapping angle is large, there is a region of the free membrane segment where the angle $\psi$ is constant, named $\psi_0$, which implies $\dot{\psi}|_{\psi_0} = 0$ and $\ddot{\psi}|_{\psi_0} = 0$. As the force $\mathscr{F}$ must be constant through the contour length at the equilibrium, its vertical and horizontal components can be written in terms of $\psi_0$ as
\begin{eqnarray}
\mathscr{F}_y&=&-(\Sigma+\omega_s)\sin \psi_0, \nonumber \\
\mathscr{F}_x&=&-(\Sigma+\omega_s)\cos \psi_0.
\end{eqnarray}
Therefore we have $\mathscr{F}_y / \mathscr{F}_x=\tan \psi_0$ and considering Eq. (\ref{f_x}), we can write the following equations
\begin{align}
\cos \psi_0=\frac{\Sigma}{\Sigma+\omega_s}; \hspace*{5mm} \tan \psi_0 = - \frac{\sqrt{\omega_s^2+2\Sigma\omega_s}}{\Sigma}.
\label{ell_f}
\end{align} 
These equations shows that increasing the wrapping angle dos not change the angle of the force (consequently the angle of the membrane) in the free part (Fig. \ref{fig_comparison}(b)). On the other hand, the angle of the membrane at the detachment of the target, $\psi_m$, is an increasing function of the wrapping angle(Eq. (\ref{el_psi_m})). The $P-F$ transition occurs when $\psi_0=\psi_m$. Therefore by substituting Eq. (\ref{ell_f}) into Eq. (\ref{el_psi_m}), we can determine the transition angle $\theta_f$ as  
\begin{equation}
\tan \theta_f=-D^2 \frac{\sqrt{\omega_s^2+2\Sigma\omega_s}}{\Sigma},
\label{el_teta_f}
\end{equation}
this equation extends the one of a cylindrical object with $D=1$\cite{khosravanizadeh2019wrapping}. The full wrapped regime occur in $\alpha=2\theta_f$. 

Now, the vertical component of the force can be written as
\begin{equation}
\mathscr{F}_y = -\Sigma \tan \psi_0 =\frac{1}{\sin \psi_0} \left[ \frac{\kappa}{2} \dot{\psi}_m^2- \Sigma ( 1 - \cos \psi_0) - \omega_s \right].
\label{f_y_1}
\end{equation}
Using Eq. (\ref{ell_f}) and some manipulation we have
\begin{equation}
2(\Sigma+\omega_s) = \frac{\kappa}{2}\dot{\psi}_m^2.
\label{el_wf_1}
\end{equation} 
By substituting $\dot{\psi}_m$ from Eq. (\ref{el_psidot_m}) and using Eq. (\ref{el_teta_f}), the membrane-target adhesion energy for partial-full wrapped transition can be found as  
\begin{equation}
\omega_f=\frac{\kappa}{2a^2}\left[ \sqrt{2\left( \bar{\Sigma}+\bar{\omega}_s \right) } + \frac{\left( \bar{\Sigma}^2+D^2(\bar{\omega}_s^2+2 \bar{\Sigma} \bar{\omega}_s) \right)^{3/2}}{D(\bar{\Sigma}+\bar{\omega}_s)^3} \right]^2,
\label{el_omega-f}
\end{equation}
where $\bar{\Sigma}$ and $\bar{\omega}_s$ are defined as $\bar{\Sigma} \equiv \frac{2 \Sigma a^2}{\kappa}$, $\bar{\omega}_s \equiv \frac{2 \omega_s a^2}{\kappa}$, respectively. This equation confirms the cylindrical equation with $D=1$\cite{khosravanizadeh2019wrapping}.
This transition is shown as a dashed line in Fig. \ref{fig3}(b).

The energy of the system in the full wrapped regime is degenerate\cite{khosravanizadeh2019wrapping}, and the left and right angles can fluctuate between $\theta_f$ and $2\pi-\theta_f$ which means the engulfment can be asymmetric (see Fig. \ref{fig_sup}). The P-F transition angle, $\alpha=2\theta_f$, (Eq. (\ref{el_teta_f})) for the oblate configuration is $\alpha\simeq \pi$, while for the prolate case it is mostly close to $2\pi$. As a result, the membrane in oblate configuration is more likely to be asymmetric than in the prolate one. Moreover, because there is an energy barrier in the highly curved tips, if the membrane could pass just one of tips it will engulf the whole particle.

\section{Discussion and Conclusion}
In this paper, we used a 2D coarse-grained model of lipid membranes to investigate the impact of local curvature of particles in the wrapping process. This model is constructed, by discretization of the Helfrich energy. Using this model we studied the engulfment of a very long cylindrical nanoparticle with elliptical cross section. The lipid membrane itself is adhered to a planar substrate mimicking the cortical cytoskeleton. To understand the role of local curvature, we have studied a system with two different orientations of the particle: in one case the object is placed upon the membrane from its long side (oblate configuration), and in the second case it is introduced to the membrane from its highly curved tip (prolate state). While the elastic parameters of the membrane are fixed, the competition between the membrane adhesion energy with the target (characterized by the parameter $\omega$) and with the cytoskeleton (parameter $\omega_s$) defines three distinct regimes of engulfment, separated by two phase transitions (see Fig. \ref{fig3}).

The target remains unwrapped (U) by the membrane if $\omega<\omega_u$, given by Eq. (\ref{el_omega-u}). Owing {to the} high curvature at the tip the values of $\omega_u$ in the prolate case is larger than the oblate configuration (Eq. (\ref{el_omega-u})). This means that in the initial stages of phagocytosis, wrapping of particles with small curvature is easier than the particles with high curvature. For $\omega > \omega_u$ the particle is partially engulfed (P) and the wrapping angle continuously grow with increasing $\omega$. In the partially wrapped regime, the prolate case shows a more rapid progression of the wrapping angle with the adhesion energy than the oblate one. The target is fully engulfed (F) by the membrane if $\omega>\omega_f$, given by Eq. (\ref{el_omega-f}). 

Both threshold target adhesion energies $\omega_u$ and $\omega_f$ are depends on the shape and the local curvature of the object, given by the aspect ratio $D$. Both $\omega_u$ and $\omega_f$ are increasing functions of the cytoskeleton adhesion energy $\omega_s$ and of the membrane bending rigidity $\kappa$. Although the full critical adhesion $\omega_f$ strongly depends on the membrane tension, the partially wrapping threshold is insensitive to membrane tension. 

It is worth mentioning  that cellular processes such as phagocytosis are strongly dependent upon actin polymerization and myosin motor proteins. While one can add these forces to the simulation\cite{atakhani2019influence,sadhu2022theoretical}, the results can be anticipated by the current model. In the prolate case, the active forces coming from the cytoskeleton can continuously grow around the particle and assist the wrapping process. On the other hand, in the oblate case, the active forces coming from the cytoskeleton must experience a strong directional change at the highly curvature tips. This directional change consumes even more energy and increase the values of $\omega_f$ for oblate case. It also should be mentioned that phagocytosis strongly depends on the diffusion time of the phagocytic receptors and if during a specific time the process can not be completed these receptors will diffuse and the particle can not be internalized\cite{swanson2008shaping,may2001phagocytosis}. Although the present model is a purely equilibrium one without kinetics, one can infer from the dependence of the wrapping angle with the adhesion energy that there exists a high energy barrier associated with wrapping the highly curved sides in the oblate case, while such barrier is absent for the prolate case. In the latter case, diffusing receptors on the cell membrane can be expected to find ligands on the particle easily, while in the former case, there is a bottleneck around the highly curvature tip which makes the process less likely. These results are in agreement with experimental results of Champion et. al \cite{champion2006role}, where they observed that the ellipsoidal disks which were introduced to neutrophils by their tips could completely be engulfed after three minutes. But, even after 110 minutes, these particles could not enter to cells when they were introduced to neutrophils by their long sides. One possible extension to bring the active forces to this model is to tune the adhesion between the membrane and the target, $\omega$, as the function of the wrapping angle and over time. This provides an extra force that can help to overcome the U-P and P-F thresholds. Such a similar method can also be used for the membrane-cytoskeleton adhesion energy, $\omega_s$, to describe the reorganization of the cytoskeleton over the wrapping process.
\section{Conflicts of Interest}
There are no conflicts to declare.
\section{Acknowledgment} 
The authors thank S.M. Hashemi for helpful discussions. And A. Kh. thanks S. Dmitrieff for his hospitality.
 \appendix
 \renewcommand\thefigure{S.\arabic{figure}} 
 \setcounter{figure}{0}  
    \section{Curvature of the ellipse}
Here we derive the equation of the curvature at the membrane-target detachment point. According to Fig. \ref{fig1}, the Cartesian components $x$ and $y$ can be written as,
\begin{align}\label{el_x_y}
\left\lbrace 
\begin{array}{l}
x =r(\theta) \sin \theta \\
y =-r(\theta) \cos \theta 
\end{array} \right..
\end{align}
By replacing Eq. \ref{el_x_y} in Eq. \ref{el_cartezian} the ellipse is described as,
\begin{equation}\label{el_polar}
\vec{r}(\theta)=\frac{ab}{\sqrt{b^2 \sin^2 \theta + a^2 \cos^2 \theta}} \, \hat{e}_r,
\end{equation}
where $\hat{e}_r$ and $\hat{e}_{\theta}$ are the radial and azimuthal unit vectors, written as 
\begin{align}\label{e_r_e_t}
\left\lbrace 
\begin{array}{c}
\hat{e}_r=\hat{i} \sin \theta -\hat{j} \cos \theta \\
\hat{e}_{\theta}=\hat{i} \cos \theta +\hat{j} \sin \theta
\end{array} \right..
\end{align}
By differentiation of $\vec{r}(\theta)$ with respect to the contour length, $s$, the tangent unit vector of the ellipse can be calculated as, 
\begin{align}\label{el_tan}
\hat{t} &=\partial_s \vec{r}=\frac{d\vec{r}}{d \theta} \frac{d \theta}{ds} \nonumber \\
&=\frac{a \sec \theta }{\sqrt{D^2 + \tan^2 \theta}} \left[ \frac{(D^2-1)\tan \theta}{(D^2 + \tan^2 \theta)}\, \hat{e}_r + \hat{e}_{\theta} \right] \frac{d \theta}{ds},
\end{align}
where $ds$ is written as,
\begin{align}\label{el_ds}
ds &= \vert \frac{d\vec{r}}{d \theta} \vert d \theta  \nonumber \\
&=\frac{a \sec \theta }{\sqrt{D^2 + \tan^2 \theta}} \left[ \frac{(D^2-1)^2\tan^2 \theta}{(D^2 + \tan^2 \theta)^2}\, + 1 \right]^{1/2} d \theta.
\end{align}
By replacing $\frac{d \theta}{ds}$ from Eq. \ref{el_ds} in Eq. \ref{el_tan} the tangent unit vector of the ellipse can be described as a function of the wrapping angle $\theta$ as
\begin{equation}\label{el_tangent}
\hat{t} =\left[ \frac{(D^2-1)^2\tan^2 \theta}{(D^2 + \tan^2 \theta)^2}\, + 1 \right]^{-1/2} \left[ \frac{(D^2-1)\tan \theta}{(D^2 + \tan^2 \theta)}\, \hat{e}_r + \hat{e}_{\theta} \right],
\end{equation}
Using Eq. \ref{e_r_e_t}, one can find the angle of the membrane at the point $m$ where it detaches from the ellipse (see Fig. \ref{fig1}(a)), 
\begin{align}\label{el_psi_m}
\cos \psi_m &=\hat{t} \cdot \hat{i}= \frac{D^2}{\sqrt{D^4+\tan^2\theta}}, \nonumber \\
\tan \psi_m &= \tan \theta /D^2.
\end{align}
where $\psi$ is the angle between the tangent vector of the membrane and the $x$ axis. The local curvature of the ellipse as a function of the wrapping angle can be found by differentiation of $\hat{t}(\theta)$ with respect to the contour length:
\begin{align}\label{el_curvature}
C(\theta) =\partial_s \hat{t}=\frac{D}{b} \left[ \frac{D^2+\tan^2\theta}{D^4+\tan^2\theta}\right]^{3/2}.
\end{align}

\section*{Supporting information}

\bibliography{citations.bib}

\begin{thebibliography}{51}%
\makeatletter
\providecommand \@ifxundefined [1]{%
 \@ifx{#1\undefined}
}%
\providecommand \@ifnum [1]{%
 \ifnum #1\expandafter \@firstoftwo
 \else \expandafter \@secondoftwo
 \fi
}%
\providecommand \@ifx [1]{%
 \ifx #1\expandafter \@firstoftwo
 \else \expandafter \@secondoftwo
 \fi
}%
\providecommand \natexlab [1]{#1}%
\providecommand \enquote  [1]{``#1''}%
\providecommand \bibnamefont  [1]{#1}%
\providecommand \bibfnamefont [1]{#1}%
\providecommand \citenamefont [1]{#1}%
\providecommand \href@noop [0]{\@secondoftwo}%
\providecommand \href [0]{\begingroup \@sanitize@url \@href}%
\providecommand \@href[1]{\@@startlink{#1}\@@href}%
\providecommand \@@href[1]{\endgroup#1\@@endlink}%
\providecommand \@sanitize@url [0]{\catcode `\\12\catcode `\$12\catcode
  `\&12\catcode `\#12\catcode `\^12\catcode `\_12\catcode `\%12\relax}%
\providecommand \@@startlink[1]{}%
\providecommand \@@endlink[0]{}%
\providecommand \url  [0]{\begingroup\@sanitize@url \@url }%
\providecommand \@url [1]{\endgroup\@href {#1}{\urlprefix }}%
\providecommand \urlprefix  [0]{URL }%
\providecommand \Eprint [0]{\href }%
\providecommand \doibase [0]{http://dx.doi.org/}%
\providecommand \selectlanguage [0]{\@gobble}%
\providecommand \bibinfo  [0]{\@secondoftwo}%
\providecommand \bibfield  [0]{\@secondoftwo}%
\providecommand \translation [1]{[#1]}%
\providecommand \BibitemOpen [0]{}%
\providecommand \bibitemStop [0]{}%
\providecommand \bibitemNoStop [0]{.\EOS\space}%
\providecommand \EOS [0]{\spacefactor3000\relax}%
\providecommand \BibitemShut  [1]{\csname bibitem#1\endcsname}%
\let\auto@bib@innerbib\@empty
\bibitem [{\citenamefont {Lodish}\ \emph {et~al.}(2000)\citenamefont {Lodish},
  \citenamefont {Berk}, \citenamefont {Zipursky}, \citenamefont {Matsudaira},
  \citenamefont {Baltimore},\ and\ \citenamefont {Darnell}}]{lodish}%
  \BibitemOpen
  \bibfield  {author} {\bibinfo {author} {\bibfnamefont {H.}~\bibnamefont
  {Lodish}}, \bibinfo {author} {\bibfnamefont {A.}~\bibnamefont {Berk}},
  \bibinfo {author} {\bibfnamefont {L.}~\bibnamefont {Zipursky}}, \bibinfo
  {author} {\bibfnamefont {P.}~\bibnamefont {Matsudaira}}, \bibinfo {author}
  {\bibfnamefont {D.}~\bibnamefont {Baltimore}}, \ and\ \bibinfo {author}
  {\bibfnamefont {J.}~\bibnamefont {Darnell}},\ }\href@noop {} {\emph {\bibinfo
  {title} {{Molecular Cell Biology}}}},\ \bibinfo {edition} {4th}\ ed.\
  (\bibinfo  {publisher} {W. H. Freeman},\ \bibinfo {year} {2000})\BibitemShut
  {NoStop}%
\bibitem [{\citenamefont {Xia}(2008)}]{xia2008nanomaterials}%
  \BibitemOpen
  \bibfield  {author} {\bibinfo {author} {\bibfnamefont {Y.}~\bibnamefont
  {Xia}},\ }\href@noop {} {\bibfield  {journal} {\bibinfo  {journal} {Nature
  Materials}\ }\textbf {\bibinfo {volume} {7}},\ \bibinfo {pages} {758}
  (\bibinfo {year} {2008})}\BibitemShut {NoStop}%
\bibitem [{\citenamefont {Weissleder}(2006)}]{weissleder2006molecular}%
  \BibitemOpen
  \bibfield  {author} {\bibinfo {author} {\bibfnamefont {R.}~\bibnamefont
  {Weissleder}},\ }\href@noop {} {\bibfield  {journal} {\bibinfo  {journal}
  {Science}\ }\textbf {\bibinfo {volume} {312}},\ \bibinfo {pages} {1168}
  (\bibinfo {year} {2006})}\BibitemShut {NoStop}%
\bibitem [{\citenamefont {Nel}\ \emph {et~al.}(2009)\citenamefont {Nel},
  \citenamefont {M{\"a}dler}, \citenamefont {Velegol}, \citenamefont {Xia},
  \citenamefont {Hoek}, \citenamefont {Somasundaran}, \citenamefont {Klaessig},
  \citenamefont {Castranova},\ and\ \citenamefont
  {Thompson}}]{nel2009understanding}%
  \BibitemOpen
  \bibfield  {author} {\bibinfo {author} {\bibfnamefont {A.~E.}\ \bibnamefont
  {Nel}}, \bibinfo {author} {\bibfnamefont {L.}~\bibnamefont {M{\"a}dler}},
  \bibinfo {author} {\bibfnamefont {D.}~\bibnamefont {Velegol}}, \bibinfo
  {author} {\bibfnamefont {T.}~\bibnamefont {Xia}}, \bibinfo {author}
  {\bibfnamefont {E.~M.~V.}\ \bibnamefont {Hoek}}, \bibinfo {author}
  {\bibfnamefont {P.}~\bibnamefont {Somasundaran}}, \bibinfo {author}
  {\bibfnamefont {F.}~\bibnamefont {Klaessig}}, \bibinfo {author}
  {\bibfnamefont {V.}~\bibnamefont {Castranova}}, \ and\ \bibinfo {author}
  {\bibfnamefont {M.}~\bibnamefont {Thompson}},\ }\href@noop {} {\bibfield
  {journal} {\bibinfo  {journal} {Nature Materials}\ }\textbf {\bibinfo
  {volume} {8}},\ \bibinfo {pages} {543} (\bibinfo {year} {2009})}\BibitemShut
  {NoStop}%
\bibitem [{\citenamefont {Allen}\ and\ \citenamefont
  {Cullis}(2004)}]{allen2004drug}%
  \BibitemOpen
  \bibfield  {author} {\bibinfo {author} {\bibfnamefont {T.~M.}\ \bibnamefont
  {Allen}}\ and\ \bibinfo {author} {\bibfnamefont {P.~R.}\ \bibnamefont
  {Cullis}},\ }\href@noop {} {\bibfield  {journal} {\bibinfo  {journal}
  {Science}\ }\textbf {\bibinfo {volume} {303}},\ \bibinfo {pages} {1818}
  (\bibinfo {year} {2004})}\BibitemShut {NoStop}%
\bibitem [{\citenamefont {Whitehead}\ \emph {et~al.}(2009)\citenamefont
  {Whitehead}, \citenamefont {Langer},\ and\ \citenamefont
  {Anderson}}]{whitehead2009knocking}%
  \BibitemOpen
  \bibfield  {author} {\bibinfo {author} {\bibfnamefont {K.~A.}\ \bibnamefont
  {Whitehead}}, \bibinfo {author} {\bibfnamefont {R.}~\bibnamefont {Langer}}, \
  and\ \bibinfo {author} {\bibfnamefont {D.~G.}\ \bibnamefont {Anderson}},\
  }\href@noop {} {\bibfield  {journal} {\bibinfo  {journal} {Nature Reviews
  Drug Discovery}\ }\textbf {\bibinfo {volume} {8}},\ \bibinfo {pages} {129}
  (\bibinfo {year} {2009})}\BibitemShut {NoStop}%
\bibitem [{\citenamefont {Peer}\ \emph {et~al.}(2007)\citenamefont {Peer},
  \citenamefont {Karp}, \citenamefont {Hong}, \citenamefont {Farokhzad},
  \citenamefont {Margalit},\ and\ \citenamefont
  {Langer}}]{peer2007nanocarriers}%
  \BibitemOpen
  \bibfield  {author} {\bibinfo {author} {\bibfnamefont {D.}~\bibnamefont
  {Peer}}, \bibinfo {author} {\bibfnamefont {J.~M.}\ \bibnamefont {Karp}},
  \bibinfo {author} {\bibfnamefont {S.}~\bibnamefont {Hong}}, \bibinfo {author}
  {\bibfnamefont {O.~C.}\ \bibnamefont {Farokhzad}}, \bibinfo {author}
  {\bibfnamefont {R.}~\bibnamefont {Margalit}}, \ and\ \bibinfo {author}
  {\bibfnamefont {R.}~\bibnamefont {Langer}},\ }\href@noop {} {\bibfield
  {journal} {\bibinfo  {journal} {Nature Nanotechnology}\ }\textbf {\bibinfo
  {volume} {2}},\ \bibinfo {pages} {751} (\bibinfo {year} {2007})}\BibitemShut
  {NoStop}%
\bibitem [{\citenamefont {Conner}\ and\ \citenamefont
  {Schmid}(2003)}]{conner2003regulated}%
  \BibitemOpen
  \bibfield  {author} {\bibinfo {author} {\bibfnamefont {S.~D.}\ \bibnamefont
  {Conner}}\ and\ \bibinfo {author} {\bibfnamefont {S.~L.}\ \bibnamefont
  {Schmid}},\ }\href@noop {} {\bibfield  {journal} {\bibinfo  {journal}
  {Nature}\ }\textbf {\bibinfo {volume} {422}},\ \bibinfo {pages} {37}
  (\bibinfo {year} {2003})}\BibitemShut {NoStop}%
\bibitem [{\citenamefont {Swanson}(2008)}]{swanson2008shaping}%
  \BibitemOpen
  \bibfield  {author} {\bibinfo {author} {\bibfnamefont {J.~A.}\ \bibnamefont
  {Swanson}},\ }\href@noop {} {\bibfield  {journal} {\bibinfo  {journal} {Nat.
  Rev. Mol. Cell Biol.}\ }\textbf {\bibinfo {volume} {9}},\ \bibinfo {pages}
  {639} (\bibinfo {year} {2008})}\BibitemShut {NoStop}%
\bibitem [{\citenamefont {Gao}\ \emph {et~al.}(2005)\citenamefont {Gao},
  \citenamefont {Shi},\ and\ \citenamefont {Freund}}]{gao2005mechanics}%
  \BibitemOpen
  \bibfield  {author} {\bibinfo {author} {\bibfnamefont {H.}~\bibnamefont
  {Gao}}, \bibinfo {author} {\bibfnamefont {W.}~\bibnamefont {Shi}}, \ and\
  \bibinfo {author} {\bibfnamefont {L.~B.}\ \bibnamefont {Freund}},\
  }\href@noop {} {\bibfield  {journal} {\bibinfo  {journal} {Proceedings of the
  National Academy of Sciences}\ }\textbf {\bibinfo {volume} {102}},\ \bibinfo
  {pages} {9469} (\bibinfo {year} {2005})}\BibitemShut {NoStop}%
\bibitem [{\citenamefont {V{\'a}cha}\ \emph {et~al.}(2011)\citenamefont
  {V{\'a}cha}, \citenamefont {Martinez-Veracoechea},\ and\ \citenamefont
  {Frenkel}}]{vacha2011receptor}%
  \BibitemOpen
  \bibfield  {author} {\bibinfo {author} {\bibfnamefont {R.}~\bibnamefont
  {V{\'a}cha}}, \bibinfo {author} {\bibfnamefont {F.~J.}\ \bibnamefont
  {Martinez-Veracoechea}}, \ and\ \bibinfo {author} {\bibfnamefont
  {D.}~\bibnamefont {Frenkel}},\ }\href@noop {} {\bibfield  {journal} {\bibinfo
   {journal} {Nano Letters}\ }\textbf {\bibinfo {volume} {11}},\ \bibinfo
  {pages} {5391} (\bibinfo {year} {2011})}\BibitemShut {NoStop}%
\bibitem [{\citenamefont {Li}\ \emph {et~al.}(2012)\citenamefont {Li},
  \citenamefont {Yue}, \citenamefont {Yang},\ and\ \citenamefont
  {Zhang}}]{li2012molecular}%
  \BibitemOpen
  \bibfield  {author} {\bibinfo {author} {\bibfnamefont {Y.}~\bibnamefont
  {Li}}, \bibinfo {author} {\bibfnamefont {T.}~\bibnamefont {Yue}}, \bibinfo
  {author} {\bibfnamefont {K.}~\bibnamefont {Yang}}, \ and\ \bibinfo {author}
  {\bibfnamefont {X.}~\bibnamefont {Zhang}},\ }\href@noop {} {\bibfield
  {journal} {\bibinfo  {journal} {Biomaterials}\ }\textbf {\bibinfo {volume}
  {33}},\ \bibinfo {pages} {4965} (\bibinfo {year} {2012})}\BibitemShut
  {NoStop}%
\bibitem [{\citenamefont {Yi}\ \emph {et~al.}(2011)\citenamefont {Yi},
  \citenamefont {Shi},\ and\ \citenamefont {Gao}}]{yi2011cellular}%
  \BibitemOpen
  \bibfield  {author} {\bibinfo {author} {\bibfnamefont {X.}~\bibnamefont
  {Yi}}, \bibinfo {author} {\bibfnamefont {X.}~\bibnamefont {Shi}}, \ and\
  \bibinfo {author} {\bibfnamefont {H.}~\bibnamefont {Gao}},\ }\href@noop {}
  {\bibfield  {journal} {\bibinfo  {journal} {Physical Review Letters}\
  }\textbf {\bibinfo {volume} {107}},\ \bibinfo {pages} {098101} (\bibinfo
  {year} {2011})}\BibitemShut {NoStop}%
\bibitem [{\citenamefont {Yuan}\ \emph {et~al.}(2010)\citenamefont {Yuan},
  \citenamefont {Li}, \citenamefont {Bao},\ and\ \citenamefont
  {Zhang}}]{yuan2010variable}%
  \BibitemOpen
  \bibfield  {author} {\bibinfo {author} {\bibfnamefont {H.}~\bibnamefont
  {Yuan}}, \bibinfo {author} {\bibfnamefont {J.}~\bibnamefont {Li}}, \bibinfo
  {author} {\bibfnamefont {G.}~\bibnamefont {Bao}}, \ and\ \bibinfo {author}
  {\bibfnamefont {S.}~\bibnamefont {Zhang}},\ }\href@noop {} {\bibfield
  {journal} {\bibinfo  {journal} {Physical Review Letters}\ }\textbf {\bibinfo
  {volume} {105}},\ \bibinfo {pages} {138101} (\bibinfo {year}
  {2010})}\BibitemShut {NoStop}%
\bibitem [{\citenamefont {Reynwar}\ \emph {et~al.}(2007)\citenamefont
  {Reynwar}, \citenamefont {Illya}, \citenamefont {Harmandaris}, \citenamefont
  {M{\"u}ller}, \citenamefont {Kremer},\ and\ \citenamefont
  {Deserno}}]{reynwar2007aggregation}%
  \BibitemOpen
  \bibfield  {author} {\bibinfo {author} {\bibfnamefont {B.~J.}\ \bibnamefont
  {Reynwar}}, \bibinfo {author} {\bibfnamefont {G.}~\bibnamefont {Illya}},
  \bibinfo {author} {\bibfnamefont {V.~A.}\ \bibnamefont {Harmandaris}},
  \bibinfo {author} {\bibfnamefont {M.~M.}\ \bibnamefont {M{\"u}ller}},
  \bibinfo {author} {\bibfnamefont {K.}~\bibnamefont {Kremer}}, \ and\ \bibinfo
  {author} {\bibfnamefont {M.}~\bibnamefont {Deserno}},\ }\href@noop {}
  {\bibfield  {journal} {\bibinfo  {journal} {Nature}\ }\textbf {\bibinfo
  {volume} {447}},\ \bibinfo {pages} {461} (\bibinfo {year}
  {2007})}\BibitemShut {NoStop}%
\bibitem [{\citenamefont {Yang}\ and\ \citenamefont
  {Ma}(2010)}]{yang2010computer}%
  \BibitemOpen
  \bibfield  {author} {\bibinfo {author} {\bibfnamefont {K.}~\bibnamefont
  {Yang}}\ and\ \bibinfo {author} {\bibfnamefont {Y.-Q.}\ \bibnamefont {Ma}},\
  }\href@noop {} {\bibfield  {journal} {\bibinfo  {journal} {Nature
  nanotechnology}\ }\textbf {\bibinfo {volume} {5}},\ \bibinfo {pages} {579}
  (\bibinfo {year} {2010})}\BibitemShut {NoStop}%
\bibitem [{\citenamefont {Deserno}(2004)}]{deserno2004elastic}%
  \BibitemOpen
  \bibfield  {author} {\bibinfo {author} {\bibfnamefont {M.}~\bibnamefont
  {Deserno}},\ }\href@noop {} {\bibfield  {journal} {\bibinfo  {journal}
  {Physical Review E}\ }\textbf {\bibinfo {volume} {69}},\ \bibinfo {pages}
  {031903} (\bibinfo {year} {2004})}\BibitemShut {NoStop}%
\bibitem [{\citenamefont {Agudo-Canalejo}(2021)}]{agudo2021particle}%
  \BibitemOpen
  \bibfield  {author} {\bibinfo {author} {\bibfnamefont {J.}~\bibnamefont
  {Agudo-Canalejo}},\ }\href@noop {} {\bibfield  {journal} {\bibinfo  {journal}
  {Soft Matter}\ }\textbf {\bibinfo {volume} {17}},\ \bibinfo {pages} {298}
  (\bibinfo {year} {2021})}\BibitemShut {NoStop}%
\bibitem [{\citenamefont {Weikl}(2003)}]{weikl2003indirect}%
  \BibitemOpen
  \bibfield  {author} {\bibinfo {author} {\bibfnamefont {T.~R.}\ \bibnamefont
  {Weikl}},\ }\href@noop {} {\bibfield  {journal} {\bibinfo  {journal} {The
  European Physical Journal E}\ }\textbf {\bibinfo {volume} {12}},\ \bibinfo
  {pages} {265} (\bibinfo {year} {2003})}\BibitemShut {NoStop}%
\bibitem [{\citenamefont {Khosravanizadeh}\ \emph {et~al.}(2019)\citenamefont
  {Khosravanizadeh}, \citenamefont {Sens},\ and\ \citenamefont
  {Mohammad-Rafiee}}]{khosravanizadeh2019wrapping}%
  \BibitemOpen
  \bibfield  {author} {\bibinfo {author} {\bibfnamefont {A.}~\bibnamefont
  {Khosravanizadeh}}, \bibinfo {author} {\bibfnamefont {P.}~\bibnamefont
  {Sens}}, \ and\ \bibinfo {author} {\bibfnamefont {F.}~\bibnamefont
  {Mohammad-Rafiee}},\ }\href@noop {} {\bibfield  {journal} {\bibinfo
  {journal} {Soft matter}\ }\textbf {\bibinfo {volume} {15}},\ \bibinfo {pages}
  {7490} (\bibinfo {year} {2019})}\BibitemShut {NoStop}%
\bibitem [{\citenamefont {Hashemi}\ \emph {et~al.}(2014)\citenamefont
  {Hashemi}, \citenamefont {Sens},\ and\ \citenamefont
  {Mohammad-Rafiee}}]{hashemi2014regulation}%
  \BibitemOpen
  \bibfield  {author} {\bibinfo {author} {\bibfnamefont {S.~M.}\ \bibnamefont
  {Hashemi}}, \bibinfo {author} {\bibfnamefont {P.}~\bibnamefont {Sens}}, \
  and\ \bibinfo {author} {\bibfnamefont {F.}~\bibnamefont {Mohammad-Rafiee}},\
  }\href@noop {} {\bibfield  {journal} {\bibinfo  {journal} {J. R. Soc.
  Interface}\ }\textbf {\bibinfo {volume} {11}},\ \bibinfo {pages} {20140769}
  (\bibinfo {year} {2014})}\BibitemShut {NoStop}%
\bibitem [{\citenamefont {Yue}\ and\ \citenamefont
  {Zhang}(2013)}]{yue2013molecular}%
  \BibitemOpen
  \bibfield  {author} {\bibinfo {author} {\bibfnamefont {T.}~\bibnamefont
  {Yue}}\ and\ \bibinfo {author} {\bibfnamefont {X.}~\bibnamefont {Zhang}},\
  }\href@noop {} {\bibfield  {journal} {\bibinfo  {journal} {Soft Matter}\
  }\textbf {\bibinfo {volume} {9}},\ \bibinfo {pages} {559} (\bibinfo {year}
  {2013})}\BibitemShut {NoStop}%
\bibitem [{\citenamefont {M{\"u}ller}\ \emph {et~al.}(2005)\citenamefont
  {M{\"u}ller}, \citenamefont {Deserno},\ and\ \citenamefont
  {Guven}}]{muller2005interface}%
  \BibitemOpen
  \bibfield  {author} {\bibinfo {author} {\bibfnamefont {M.~M.}\ \bibnamefont
  {M{\"u}ller}}, \bibinfo {author} {\bibfnamefont {M.}~\bibnamefont {Deserno}},
  \ and\ \bibinfo {author} {\bibfnamefont {J.}~\bibnamefont {Guven}},\
  }\href@noop {} {\bibfield  {journal} {\bibinfo  {journal} {Physical Review
  E}\ }\textbf {\bibinfo {volume} {72}},\ \bibinfo {pages} {061407} (\bibinfo
  {year} {2005})}\BibitemShut {NoStop}%
\bibitem [{\citenamefont {Irmscher}\ \emph {et~al.}(2013)\citenamefont
  {Irmscher}, \citenamefont {de~Jong}, \citenamefont {Kress},\ and\
  \citenamefont {Prins}}]{irmscher2013method}%
  \BibitemOpen
  \bibfield  {author} {\bibinfo {author} {\bibfnamefont {M.}~\bibnamefont
  {Irmscher}}, \bibinfo {author} {\bibfnamefont {A.~M.}\ \bibnamefont
  {de~Jong}}, \bibinfo {author} {\bibfnamefont {H.}~\bibnamefont {Kress}}, \
  and\ \bibinfo {author} {\bibfnamefont {M.~W.}\ \bibnamefont {Prins}},\
  }\href@noop {} {\bibfield  {journal} {\bibinfo  {journal} {Journal of The
  Royal Society Interface}\ }\textbf {\bibinfo {volume} {10}},\ \bibinfo
  {pages} {20121048} (\bibinfo {year} {2013})}\BibitemShut {NoStop}%
\bibitem [{\citenamefont {Mirigian}\ and\ \citenamefont
  {Muthukumar}(2013)}]{mirigian2013kinetics}%
  \BibitemOpen
  \bibfield  {author} {\bibinfo {author} {\bibfnamefont {S.}~\bibnamefont
  {Mirigian}}\ and\ \bibinfo {author} {\bibfnamefont {M.}~\bibnamefont
  {Muthukumar}},\ }\href@noop {} {\bibfield  {journal} {\bibinfo  {journal}
  {The Journal of chemical physics}\ }\textbf {\bibinfo {volume} {139}},\
  \bibinfo {pages} {044908} (\bibinfo {year} {2013})}\BibitemShut {NoStop}%
\bibitem [{\citenamefont {M{\"u}ller}\ \emph {et~al.}(2007)\citenamefont
  {M{\"u}ller}, \citenamefont {Deserno},\ and\ \citenamefont
  {Guven}}]{muller2007balancing}%
  \BibitemOpen
  \bibfield  {author} {\bibinfo {author} {\bibfnamefont {M.~M.}\ \bibnamefont
  {M{\"u}ller}}, \bibinfo {author} {\bibfnamefont {M.}~\bibnamefont {Deserno}},
  \ and\ \bibinfo {author} {\bibfnamefont {J.}~\bibnamefont {Guven}},\
  }\href@noop {} {\bibfield  {journal} {\bibinfo  {journal} {Physical Review
  E}\ }\textbf {\bibinfo {volume} {76}},\ \bibinfo {pages} {011921} (\bibinfo
  {year} {2007})}\BibitemShut {NoStop}%
\bibitem [{\citenamefont {Clarke}\ \emph {et~al.}(2010)\citenamefont {Clarke},
  \citenamefont {Engel}, \citenamefont {Giorgione}, \citenamefont
  {M{\"u}ller-Taubenberger}, \citenamefont {Prassler}, \citenamefont
  {Veltman},\ and\ \citenamefont {Gerisch}}]{clarke2010curvature}%
  \BibitemOpen
  \bibfield  {author} {\bibinfo {author} {\bibfnamefont {M.}~\bibnamefont
  {Clarke}}, \bibinfo {author} {\bibfnamefont {U.}~\bibnamefont {Engel}},
  \bibinfo {author} {\bibfnamefont {J.}~\bibnamefont {Giorgione}}, \bibinfo
  {author} {\bibfnamefont {A.}~\bibnamefont {M{\"u}ller-Taubenberger}},
  \bibinfo {author} {\bibfnamefont {J.}~\bibnamefont {Prassler}}, \bibinfo
  {author} {\bibfnamefont {D.}~\bibnamefont {Veltman}}, \ and\ \bibinfo
  {author} {\bibfnamefont {G.}~\bibnamefont {Gerisch}},\ }\href@noop {}
  {\bibfield  {journal} {\bibinfo  {journal} {BMC biology}\ }\textbf {\bibinfo
  {volume} {8}},\ \bibinfo {pages} {1} (\bibinfo {year} {2010})}\BibitemShut
  {NoStop}%
\bibitem [{\citenamefont {Rittig}\ \emph {et~al.}(1992)\citenamefont {Rittig},
  \citenamefont {Krause}, \citenamefont {H{\"a}upl}, \citenamefont {Schaible},
  \citenamefont {Modolell}, \citenamefont {Kramer}, \citenamefont
  {L{\"u}tjen-Drecoll}, \citenamefont {Simon},\ and\ \citenamefont
  {Burmester}}]{rittig1992coiling}%
  \BibitemOpen
  \bibfield  {author} {\bibinfo {author} {\bibfnamefont {M.}~\bibnamefont
  {Rittig}}, \bibinfo {author} {\bibfnamefont {A.}~\bibnamefont {Krause}},
  \bibinfo {author} {\bibfnamefont {T.}~\bibnamefont {H{\"a}upl}}, \bibinfo
  {author} {\bibfnamefont {U.}~\bibnamefont {Schaible}}, \bibinfo {author}
  {\bibfnamefont {M.}~\bibnamefont {Modolell}}, \bibinfo {author}
  {\bibfnamefont {M.}~\bibnamefont {Kramer}}, \bibinfo {author} {\bibfnamefont
  {E.}~\bibnamefont {L{\"u}tjen-Drecoll}}, \bibinfo {author} {\bibfnamefont
  {M.}~\bibnamefont {Simon}}, \ and\ \bibinfo {author} {\bibfnamefont
  {G.}~\bibnamefont {Burmester}},\ }\href@noop {} {\bibfield  {journal}
  {\bibinfo  {journal} {Infection and Immunity}\ }\textbf {\bibinfo {volume}
  {60}},\ \bibinfo {pages} {4205} (\bibinfo {year} {1992})}\BibitemShut
  {NoStop}%
\bibitem [{\citenamefont {Horwitz}(1984)}]{horwitz1984phagocytosis}%
  \BibitemOpen
  \bibfield  {author} {\bibinfo {author} {\bibfnamefont {M.~A.}\ \bibnamefont
  {Horwitz}},\ }\href@noop {} {\bibfield  {journal} {\bibinfo  {journal}
  {Cell}\ }\textbf {\bibinfo {volume} {36}},\ \bibinfo {pages} {27} (\bibinfo
  {year} {1984})}\BibitemShut {NoStop}%
\bibitem [{\citenamefont {Cureton}\ \emph {et~al.}(2009)\citenamefont
  {Cureton}, \citenamefont {Massol}, \citenamefont {Saffarian}, \citenamefont
  {Kirchhausen},\ and\ \citenamefont {Whelan}}]{cureton2009vesicular}%
  \BibitemOpen
  \bibfield  {author} {\bibinfo {author} {\bibfnamefont {D.~K.}\ \bibnamefont
  {Cureton}}, \bibinfo {author} {\bibfnamefont {R.~H.}\ \bibnamefont {Massol}},
  \bibinfo {author} {\bibfnamefont {S.}~\bibnamefont {Saffarian}}, \bibinfo
  {author} {\bibfnamefont {T.~L.}\ \bibnamefont {Kirchhausen}}, \ and\ \bibinfo
  {author} {\bibfnamefont {S.~P.}\ \bibnamefont {Whelan}},\ }\href@noop {}
  {\bibfield  {journal} {\bibinfo  {journal} {PLoS pathogens}\ }\textbf
  {\bibinfo {volume} {5}},\ \bibinfo {pages} {e1000394} (\bibinfo {year}
  {2009})}\BibitemShut {NoStop}%
\bibitem [{\citenamefont {Ge}\ \emph {et~al.}(2010)\citenamefont {Ge},
  \citenamefont {Tsao}, \citenamefont {Schein}, \citenamefont {Green},
  \citenamefont {Luo},\ and\ \citenamefont {Zhou}}]{ge2010cryo}%
  \BibitemOpen
  \bibfield  {author} {\bibinfo {author} {\bibfnamefont {P.}~\bibnamefont
  {Ge}}, \bibinfo {author} {\bibfnamefont {J.}~\bibnamefont {Tsao}}, \bibinfo
  {author} {\bibfnamefont {S.}~\bibnamefont {Schein}}, \bibinfo {author}
  {\bibfnamefont {T.~J.}\ \bibnamefont {Green}}, \bibinfo {author}
  {\bibfnamefont {M.}~\bibnamefont {Luo}}, \ and\ \bibinfo {author}
  {\bibfnamefont {Z.~H.}\ \bibnamefont {Zhou}},\ }\href@noop {} {\bibfield
  {journal} {\bibinfo  {journal} {Science}\ }\textbf {\bibinfo {volume}
  {327}},\ \bibinfo {pages} {689} (\bibinfo {year} {2010})}\BibitemShut
  {NoStop}%
\bibitem [{\citenamefont {Champion}\ and\ \citenamefont
  {Mitragotri}(2006)}]{champion2006role}%
  \BibitemOpen
  \bibfield  {author} {\bibinfo {author} {\bibfnamefont {J.~A.}\ \bibnamefont
  {Champion}}\ and\ \bibinfo {author} {\bibfnamefont {S.}~\bibnamefont
  {Mitragotri}},\ }\href@noop {} {\bibfield  {journal} {\bibinfo  {journal}
  {Proceedings of the National Academy of Sciences}\ }\textbf {\bibinfo
  {volume} {103}},\ \bibinfo {pages} {4930} (\bibinfo {year}
  {2006})}\BibitemShut {NoStop}%
\bibitem [{\citenamefont {Champion}\ and\ \citenamefont
  {Mitragotri}(2009)}]{champion2009shape}%
  \BibitemOpen
  \bibfield  {author} {\bibinfo {author} {\bibfnamefont {J.~A.}\ \bibnamefont
  {Champion}}\ and\ \bibinfo {author} {\bibfnamefont {S.}~\bibnamefont
  {Mitragotri}},\ }\href@noop {} {\bibfield  {journal} {\bibinfo  {journal}
  {Pharmaceutical research}\ }\textbf {\bibinfo {volume} {26}},\ \bibinfo
  {pages} {244} (\bibinfo {year} {2009})}\BibitemShut {NoStop}%
\bibitem [{\citenamefont {Gratton}\ \emph {et~al.}(2008)\citenamefont
  {Gratton}, \citenamefont {Ropp}, \citenamefont {Pohlhaus}, \citenamefont
  {Luft}, \citenamefont {Madden}, \citenamefont {Napier},\ and\ \citenamefont
  {DeSimone}}]{gratton2008effect}%
  \BibitemOpen
  \bibfield  {author} {\bibinfo {author} {\bibfnamefont {S.~E.}\ \bibnamefont
  {Gratton}}, \bibinfo {author} {\bibfnamefont {P.~A.}\ \bibnamefont {Ropp}},
  \bibinfo {author} {\bibfnamefont {P.~D.}\ \bibnamefont {Pohlhaus}}, \bibinfo
  {author} {\bibfnamefont {J.~C.}\ \bibnamefont {Luft}}, \bibinfo {author}
  {\bibfnamefont {V.~J.}\ \bibnamefont {Madden}}, \bibinfo {author}
  {\bibfnamefont {M.~E.}\ \bibnamefont {Napier}}, \ and\ \bibinfo {author}
  {\bibfnamefont {J.~M.}\ \bibnamefont {DeSimone}},\ }\href@noop {} {\bibfield
  {journal} {\bibinfo  {journal} {Proceedings of the National Academy of
  Sciences}\ }\textbf {\bibinfo {volume} {105}},\ \bibinfo {pages} {11613}
  (\bibinfo {year} {2008})}\BibitemShut {NoStop}%
\bibitem [{\citenamefont {Chen}\ \emph {et~al.}(2016)\citenamefont {Chen},
  \citenamefont {Xiao}, \citenamefont {Zhu}, \citenamefont {Wang},\ and\
  \citenamefont {Liang}}]{chen2016shape}%
  \BibitemOpen
  \bibfield  {author} {\bibinfo {author} {\bibfnamefont {L.}~\bibnamefont
  {Chen}}, \bibinfo {author} {\bibfnamefont {S.}~\bibnamefont {Xiao}}, \bibinfo
  {author} {\bibfnamefont {H.}~\bibnamefont {Zhu}}, \bibinfo {author}
  {\bibfnamefont {L.}~\bibnamefont {Wang}}, \ and\ \bibinfo {author}
  {\bibfnamefont {H.}~\bibnamefont {Liang}},\ }\href@noop {} {\bibfield
  {journal} {\bibinfo  {journal} {Soft Matter}\ }\textbf {\bibinfo {volume}
  {12}},\ \bibinfo {pages} {2632} (\bibinfo {year} {2016})}\BibitemShut
  {NoStop}%
\bibitem [{\citenamefont {Shi}\ \emph {et~al.}(2011)\citenamefont {Shi},
  \citenamefont {von Dem~Bussche}, \citenamefont {Hurt}, \citenamefont {Kane},\
  and\ \citenamefont {Gao}}]{shi2011cell}%
  \BibitemOpen
  \bibfield  {author} {\bibinfo {author} {\bibfnamefont {X.}~\bibnamefont
  {Shi}}, \bibinfo {author} {\bibfnamefont {A.}~\bibnamefont {von
  Dem~Bussche}}, \bibinfo {author} {\bibfnamefont {R.~H.}\ \bibnamefont
  {Hurt}}, \bibinfo {author} {\bibfnamefont {A.~B.}\ \bibnamefont {Kane}}, \
  and\ \bibinfo {author} {\bibfnamefont {H.}~\bibnamefont {Gao}},\ }\href@noop
  {} {\bibfield  {journal} {\bibinfo  {journal} {Nature nanotechnology}\
  }\textbf {\bibinfo {volume} {6}},\ \bibinfo {pages} {714} (\bibinfo {year}
  {2011})}\BibitemShut {NoStop}%
\bibitem [{\citenamefont {Chithrani}\ \emph {et~al.}(2006)\citenamefont
  {Chithrani}, \citenamefont {Ghazani},\ and\ \citenamefont
  {Chan}}]{chithrani2006determining}%
  \BibitemOpen
  \bibfield  {author} {\bibinfo {author} {\bibfnamefont {B.~D.}\ \bibnamefont
  {Chithrani}}, \bibinfo {author} {\bibfnamefont {A.~A.}\ \bibnamefont
  {Ghazani}}, \ and\ \bibinfo {author} {\bibfnamefont {W.~C.}\ \bibnamefont
  {Chan}},\ }\href@noop {} {\bibfield  {journal} {\bibinfo  {journal} {Nano
  letters}\ }\textbf {\bibinfo {volume} {6}},\ \bibinfo {pages} {662} (\bibinfo
  {year} {2006})}\BibitemShut {NoStop}%
\bibitem [{\citenamefont {Dasgupta}\ \emph {et~al.}(2014)\citenamefont
  {Dasgupta}, \citenamefont {Auth},\ and\ \citenamefont
  {Gompper}}]{dasgupta2014shape}%
  \BibitemOpen
  \bibfield  {author} {\bibinfo {author} {\bibfnamefont {S.}~\bibnamefont
  {Dasgupta}}, \bibinfo {author} {\bibfnamefont {T.}~\bibnamefont {Auth}}, \
  and\ \bibinfo {author} {\bibfnamefont {G.}~\bibnamefont {Gompper}},\
  }\href@noop {} {\bibfield  {journal} {\bibinfo  {journal} {Nano letters}\
  }\textbf {\bibinfo {volume} {14}},\ \bibinfo {pages} {687} (\bibinfo {year}
  {2014})}\BibitemShut {NoStop}%
\bibitem [{\citenamefont {Dasgupta}\ \emph {et~al.}(2013)\citenamefont
  {Dasgupta}, \citenamefont {Auth},\ and\ \citenamefont
  {Gompper}}]{dasgupta2013wrapping}%
  \BibitemOpen
  \bibfield  {author} {\bibinfo {author} {\bibfnamefont {S.}~\bibnamefont
  {Dasgupta}}, \bibinfo {author} {\bibfnamefont {T.}~\bibnamefont {Auth}}, \
  and\ \bibinfo {author} {\bibfnamefont {G.}~\bibnamefont {Gompper}},\
  }\href@noop {} {\bibfield  {journal} {\bibinfo  {journal} {Soft Matter}\
  }\textbf {\bibinfo {volume} {9}},\ \bibinfo {pages} {5473} (\bibinfo {year}
  {2013})}\BibitemShut {NoStop}%
\bibitem [{\citenamefont {Richards}\ and\ \citenamefont
  {Endres}(2016)}]{richards2016target}%
  \BibitemOpen
  \bibfield  {author} {\bibinfo {author} {\bibfnamefont {D.~M.}\ \bibnamefont
  {Richards}}\ and\ \bibinfo {author} {\bibfnamefont {R.~G.}\ \bibnamefont
  {Endres}},\ }\href@noop {} {\bibfield  {journal} {\bibinfo  {journal}
  {Proceedings of the National Academy of Sciences}\ }\textbf {\bibinfo
  {volume} {113}},\ \bibinfo {pages} {6113} (\bibinfo {year}
  {2016})}\BibitemShut {NoStop}%
\bibitem [{\citenamefont {Bahrami}(2013)}]{bahrami2013orientational}%
  \BibitemOpen
  \bibfield  {author} {\bibinfo {author} {\bibfnamefont {A.~H.}\ \bibnamefont
  {Bahrami}},\ }\href@noop {} {\bibfield  {journal} {\bibinfo  {journal} {Soft
  Matter}\ }\textbf {\bibinfo {volume} {9}},\ \bibinfo {pages} {8642} (\bibinfo
  {year} {2013})}\BibitemShut {NoStop}%
\bibitem [{\citenamefont {Baranov}\ \emph {et~al.}(2021)\citenamefont
  {Baranov}, \citenamefont {Kumar}, \citenamefont {Sacanna}, \citenamefont
  {Thutupalli},\ and\ \citenamefont {Van~den Bogaart}}]{baranov2021modulation}%
  \BibitemOpen
  \bibfield  {author} {\bibinfo {author} {\bibfnamefont {M.~V.}\ \bibnamefont
  {Baranov}}, \bibinfo {author} {\bibfnamefont {M.}~\bibnamefont {Kumar}},
  \bibinfo {author} {\bibfnamefont {S.}~\bibnamefont {Sacanna}}, \bibinfo
  {author} {\bibfnamefont {S.}~\bibnamefont {Thutupalli}}, \ and\ \bibinfo
  {author} {\bibfnamefont {G.}~\bibnamefont {Van~den Bogaart}},\ }\href@noop {}
  {\bibfield  {journal} {\bibinfo  {journal} {Frontiers in immunology}\ ,\
  \bibinfo {pages} {3854}} (\bibinfo {year} {2021})}\BibitemShut {NoStop}%
\bibitem [{\citenamefont {Chithrani}\ and\ \citenamefont
  {Chan}(2007)}]{chithrani2007elucidating}%
  \BibitemOpen
  \bibfield  {author} {\bibinfo {author} {\bibfnamefont {B.~D.}\ \bibnamefont
  {Chithrani}}\ and\ \bibinfo {author} {\bibfnamefont {W.~C.}\ \bibnamefont
  {Chan}},\ }\href@noop {} {\bibfield  {journal} {\bibinfo  {journal} {Nano
  letters}\ }\textbf {\bibinfo {volume} {7}},\ \bibinfo {pages} {1542}
  (\bibinfo {year} {2007})}\BibitemShut {NoStop}%
\bibitem [{\citenamefont {Huang}\ \emph {et~al.}(2013)\citenamefont {Huang},
  \citenamefont {Zhang}, \citenamefont {Yuan}, \citenamefont {Gao},\ and\
  \citenamefont {Zhang}}]{huang2013role}%
  \BibitemOpen
  \bibfield  {author} {\bibinfo {author} {\bibfnamefont {C.}~\bibnamefont
  {Huang}}, \bibinfo {author} {\bibfnamefont {Y.}~\bibnamefont {Zhang}},
  \bibinfo {author} {\bibfnamefont {H.}~\bibnamefont {Yuan}}, \bibinfo {author}
  {\bibfnamefont {H.}~\bibnamefont {Gao}}, \ and\ \bibinfo {author}
  {\bibfnamefont {S.}~\bibnamefont {Zhang}},\ }\href@noop {} {\bibfield
  {journal} {\bibinfo  {journal} {Nano letters}\ }\textbf {\bibinfo {volume}
  {13}},\ \bibinfo {pages} {4546} (\bibinfo {year} {2013})}\BibitemShut
  {NoStop}%
\bibitem [{\citenamefont {Bray}(2000)}]{bray}%
  \BibitemOpen
  \bibfield  {author} {\bibinfo {author} {\bibfnamefont {D.}~\bibnamefont
  {Bray}},\ }\href@noop {} {\emph {\bibinfo {title} {{Cell Movements: From
  Molecules to Motility}}}},\ \bibinfo {edition} {2nd}\ ed.\ (\bibinfo
  {publisher} {Garland Science},\ \bibinfo {year} {2000})\BibitemShut {NoStop}%
\bibitem [{\citenamefont {Helfrich}(1973)}]{helfrich1973elastic}%
  \BibitemOpen
  \bibfield  {author} {\bibinfo {author} {\bibfnamefont {W.}~\bibnamefont
  {Helfrich}},\ }\href@noop {} {\bibfield  {journal} {\bibinfo  {journal} {Z.
  Naturforsch. C}\ }\textbf {\bibinfo {volume} {28}},\ \bibinfo {pages} {693}
  (\bibinfo {year} {1973})}\BibitemShut {NoStop}%
\bibitem [{esp()}]{espresso}%
  \BibitemOpen
  \href@noop {} {}\bibinfo {note} {\url{http:www.espresso.mpg.de}}\BibitemShut
  {NoStop}%
\bibitem [{\citenamefont {Landau}\ and\ \citenamefont
  {Lifshitz}(1986)}]{Landau_Elasticity}%
  \BibitemOpen
  \bibfield  {author} {\bibinfo {author} {\bibfnamefont {L.}~\bibnamefont
  {Landau}}\ and\ \bibinfo {author} {\bibfnamefont {E.}~\bibnamefont
  {Lifshitz}},\ }\href@noop {} {\emph {\bibinfo {title} {Theory of
  Elasticity}}}\ (\bibinfo  {publisher} {Pergamon Press},\ \bibinfo {year}
  {1986})\BibitemShut {NoStop}%
\bibitem [{\citenamefont {Atakhani}\ \emph {et~al.}(2019)\citenamefont
  {Atakhani}, \citenamefont {Mohammad-Rafiee},\ and\ \citenamefont
  {Gholami}}]{atakhani2019influence}%
  \BibitemOpen
  \bibfield  {author} {\bibinfo {author} {\bibfnamefont {A.}~\bibnamefont
  {Atakhani}}, \bibinfo {author} {\bibfnamefont {F.}~\bibnamefont
  {Mohammad-Rafiee}}, \ and\ \bibinfo {author} {\bibfnamefont {A.}~\bibnamefont
  {Gholami}},\ }\href@noop {} {\bibfield  {journal} {\bibinfo  {journal} {PloS
  one}\ }\textbf {\bibinfo {volume} {14}},\ \bibinfo {pages} {e0213810}
  (\bibinfo {year} {2019})}\BibitemShut {NoStop}%
\bibitem [{\citenamefont {Sadhu}\ \emph {et~al.}(2022)\citenamefont {Sadhu},
  \citenamefont {Barger}, \citenamefont {Peni{\v{c}}}, \citenamefont
  {Igli{\v{c}}}, \citenamefont {Krendel}, \citenamefont {Gauthier},\ and\
  \citenamefont {Gov}}]{sadhu2022theoretical}%
  \BibitemOpen
  \bibfield  {author} {\bibinfo {author} {\bibfnamefont {R.~K.}\ \bibnamefont
  {Sadhu}}, \bibinfo {author} {\bibfnamefont {S.~R.}\ \bibnamefont {Barger}},
  \bibinfo {author} {\bibfnamefont {S.}~\bibnamefont {Peni{\v{c}}}}, \bibinfo
  {author} {\bibfnamefont {A.}~\bibnamefont {Igli{\v{c}}}}, \bibinfo {author}
  {\bibfnamefont {M.}~\bibnamefont {Krendel}}, \bibinfo {author} {\bibfnamefont
  {N.~C.}\ \bibnamefont {Gauthier}}, \ and\ \bibinfo {author} {\bibfnamefont
  {N.~S.}\ \bibnamefont {Gov}},\ }\href@noop {} {\bibfield  {journal} {\bibinfo
   {journal} {arXiv preprint arXiv:2201.01133}\ } (\bibinfo {year}
  {2022})}\BibitemShut {NoStop}%
\bibitem [{\citenamefont {May}\ and\ \citenamefont
  {Machesky}(2001)}]{may2001phagocytosis}%
  \BibitemOpen
  \bibfield  {author} {\bibinfo {author} {\bibfnamefont {R.~C.}\ \bibnamefont
  {May}}\ and\ \bibinfo {author} {\bibfnamefont {L.~M.}\ \bibnamefont
  {Machesky}},\ }\href@noop {} {\bibfield  {journal} {\bibinfo  {journal}
  {Journal of cell science}\ }\textbf {\bibinfo {volume} {114}},\ \bibinfo
  {pages} {1061} (\bibinfo {year} {2001})}\BibitemShut {NoStop}%
\end{thebibliography}%

\begin{figure*}
\centering
\includegraphics[width=15cm]{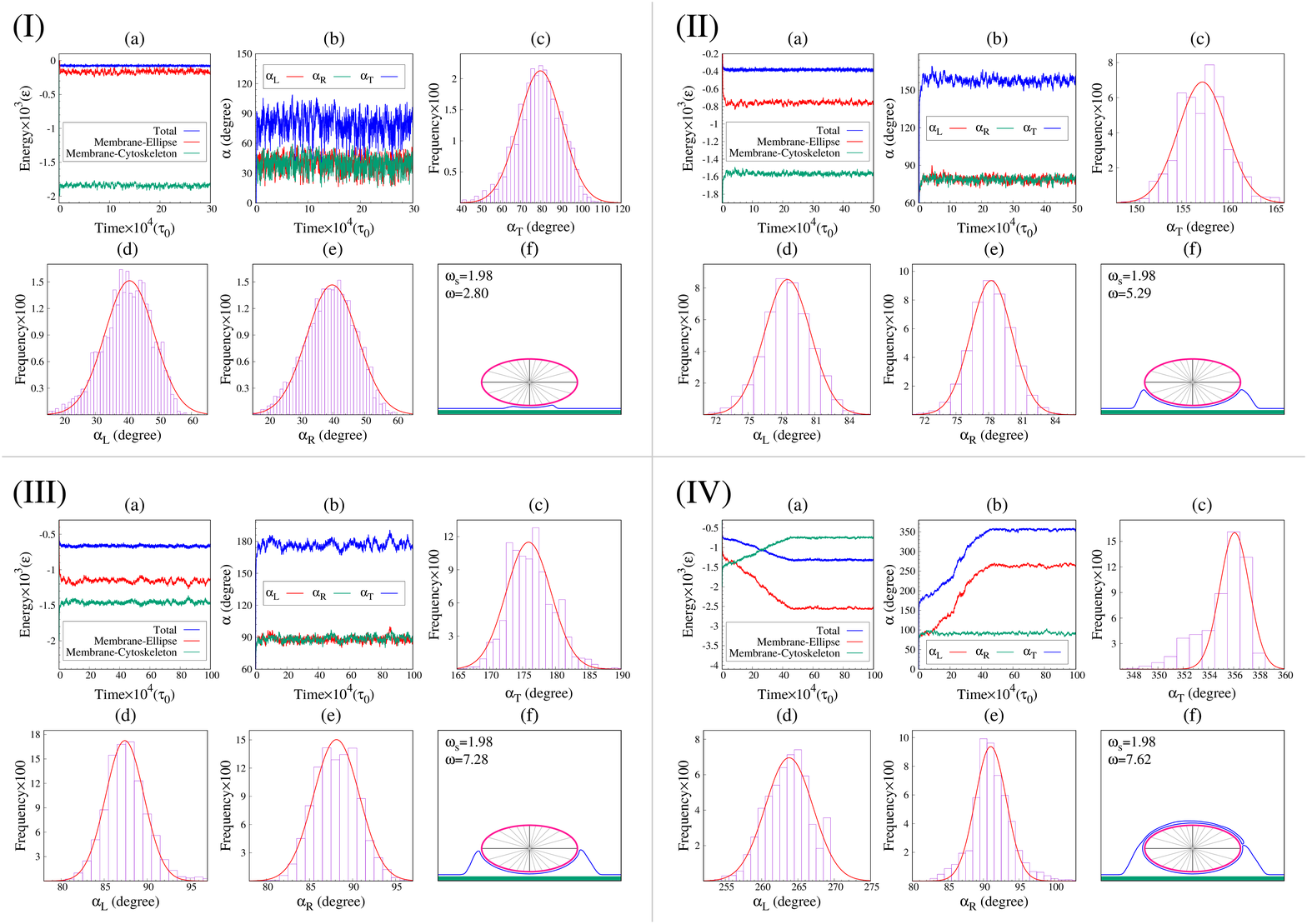}\\
\hrulefill\vspace{15pt}\par
\includegraphics[width=15cm]{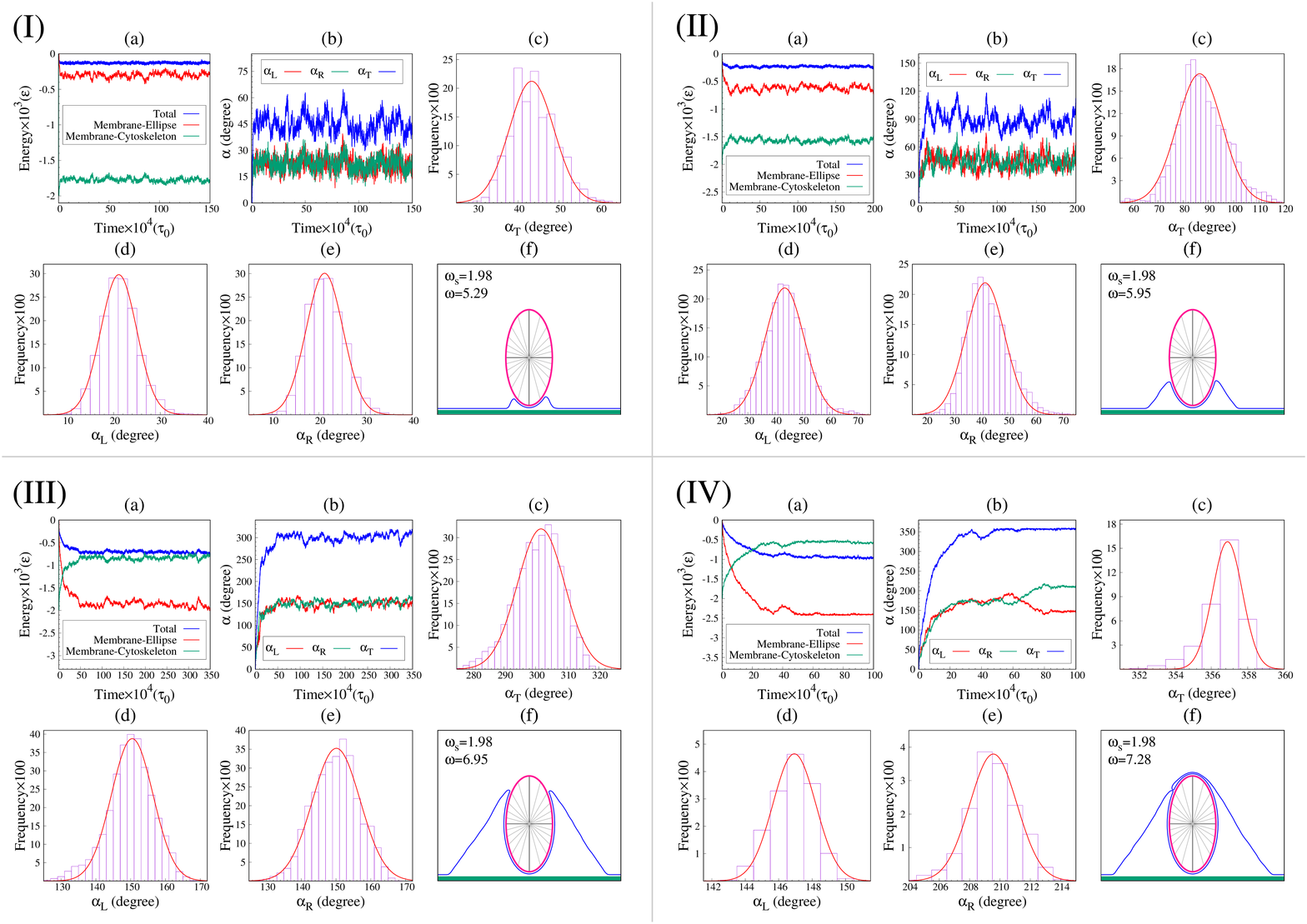}
\caption{Simulation results of oblate (top) and prolate (bottom) configurations, for different values of the membrane-target adhesion energy, $\omega$, with $\omega_s=1.98 \varepsilon/\sigma^2$. In the oblate configuration, the values of membrane-target adhesion energy from panels I to IV are $\omega=2.80, 5.29, 7.28, 7.62 \varepsilon/\sigma^2$, respectively, and for the prolate case are $\omega=5.29, 5.95, 6.95, 7.28 \varepsilon/\sigma^2$. Time evolution of (a) the total energy (blue), membrane-cytoskeleton adhesion energy (green) and membrane-target adhesion energy (red); 
(b) the total (blue), left (red) and right wrapping angles (green). Panels (c) to (e) represent the distribution of the total, the left, and the right wrapping angles in the steady-state, respectively. Panel (f) shows a typical snapshot of the membrane conformation in the equilibrium state.}
\label{fig_sup}
\end{figure*}

\end{document}